\documentclass{rmaa}

\usepackage{paralist}
\usepackage{psfrag,color}
\usepackage[latin1]{inputenc}

\title{The nature and origin of Narrow Line AGN activity in a sample of isolated SDSS galaxies.}

\author{
  R. Coziol,\altaffilmark{1}
  J.~P. Torres-Papaqui,\altaffilmark{1}
  I. Plauchu-Frayn,\altaffilmark{2}
  J.~M. Islas-Islas,\altaffilmark{1}
  R.~A. Ortega-Minakata,\altaffilmark{1}
  D.~M. Neri-Larios,\altaffilmark{1}
  and H. Andernach\altaffilmark{1}}

\altaffiltext{1}{Departamento de Astronom\'{\i}a, Universidad de
Guanajuato, Gto, M\'exico}

\altaffiltext{2}{Instituto de Astrof\'{\i}sica de Andaluc\'{\i}a
(CSIC), Granada, Spain}

\shortauthor{Coziol et al.}

\shorttitle{NLAGN in isolated galaxies}

\fulladdresses{
\item H. Andernach, R. Coziol, J.~M. Islas-Islas, D.~M. Neri-Larios, R.~A.
Ortega-Minakata, J.~P. Torres-Papaqui: Departamento de
Astronom\'{\i}a, Universidad de Guanajuato, Apartado Postal 144,
36000 Guanajuato, Gto, Mexico
  (heinz, coziol, jmislas, daniel, rene, papaqui@astro.ugto.mx).
\item
I. Plauchu-Frayn: Instituto de Astrof\'{\i}sica de Andaluc\'{\i}a
(CSIC), E-18008, Granada, Spain
  (ilse@iaa.ee).}

\listofauthors{R. Coziol, J.~P. Torres-Papaqui, I. Plauchu-Frayn,
J.~M. Islas-Islas, R.~A. Ortega-Minakata,\altaffilmark, D.~M.
Neri-Larios \& H. Andernach}
  \indexauthor{Coziol, R.}
  \indexauthor{Torres-Papaqui, J.~P.}
  \indexauthor{Plauchu-Frayn, I.}
  \indexauthor{Islas-Islas, J.~M.}
  \indexauthor{Ortega-Minakata, R.~A.}
  \indexauthor{Neri-Larios, D.~M.}
  \indexauthor{Andernach, H.}

\abstract{We discuss the nature and origin of the nuclear activity
observed in a sample of 292 SDSS narrow-emission-line galaxies,
considered to have formed and evolved in isolation. All these
galaxies are spiral like and show some kind of nuclear activity. The
fraction of Narrow Line AGNs (NLAGNs) and Transition type Objects
(TOs; a NLAGN with circumnuclear star formation) is relatively high,
amounting to 64\% of the galaxies. There is a definite trend for the
NLAGNs to appear in early-type spirals, while the star forming
galaxies and TOs are found in later-type spirals. We verify that the
probability for a galaxy to show an AGN characteristic increases
with the bulge mass of the galaxy \citep{TP11}, and find evidence
that this trend is really a by-product of the morphology, suggesting
that the AGN phenomenon is intimately connected with the formation
process of the galaxies. Consistent with this interpretation, we
establish a strong connection between the astration rate--the
efficiency with which the gas is transformed into stars--the AGN
phenomenon, and the gravitational binding energy of the galaxies:
the higher the binding energy, the higher the astration rate and the
higher the probability to find an AGN. The NLAGNs in our sample are
consistent with scaled-down or powered-down versions of quasars and
Broad Line AGNs.}

\resumen{ Discutimos la naturaleza y origen de la actividad nuclear
observada en una muestra de 292 galaxias con l\'{\i}neas de
emisi\'on angostas del SDSS, en las cuales se considera que su
formaci\'on y evoluci\'on ha ocurrido en aislamiento. Todas las
galaxias de la muestra son del tipo espiral y muestran alg\'un tipo
de actividad nuclear. La fracci\'on de galaxias con un nucleo activo
(AGNs) o de objetos de transici\'on (TOs; un AGN con formaci\'on
estelar circumnuclear) es relativamente alta, ascendiendo a un 64\%
de la muestra. Existe una clara tendencia en la cual los AGNs
tienden a encontrarse en galaxias del tipo espiral temprano,
mientras que las galaxias formando estrellas y los TOs tienden a
hacerlo en espirales tard\'{\i}as. Verificamos que la probabilidad
de que una galaxia muestre un nucleo activo aumenta con la masa de
su bulbo (Torres-Papaqui et al. 2011), tambi\'en encontramos
evidencia de que dicha tendencia es realmente un subproducto de la
morfolog\'{\i}a, sugiriendo que el fen\'omeno AGN est\'a
\'{\i}ntimamente ligado al proceso de formaci\'on de galaxias.
Consistente con esta interpretaci\'on, establecemos una conexi\'on
fuerte entre la taja de astraci\'on--la eficiencia con la cual el
gas es transformado en estrellas--el fen\'omeno AGN y la
energ\'{\i}a de ligado gravitacional de las galaxias: a mayor
energ\'{\i}a de ligado, mayor la taja de astraci\'on y mayor la
probabilidad de encontrar un AGN. Los AGNs con l\'{\i}neas de
emisi\'on angostas en nuestra muestra son consistentes con una
versi\'on a menor escala o menor energ\'{\i}a de cu\'asares y AGNs
con l\'{\i}neas de emisi\'on anchas. }

\addkeyword{galaxies: active}
\addkeyword{galaxies: formation}

\begin{document}
\maketitle

\section{Introduction}
\label{sec:intro}

Spectroscopic surveys, like the Sloan Digital Sky Survey (SDSS)
\citep{york00,stoughton02}, have revealed that a high fraction of
galaxies in the nearby universe shows emission lines consistent with
some kind of nuclear activity. Using diagnostic diagrams that
compare various emission line ratios, different classification
criteria were proposed to classify these galaxies based on the
possible sources of excitation of the gas
\citep{baldwin81,veilleux87,kewley01,kauffmann03}. These
classification criteria imply there are two possible main sources:
thermal sources, which are associated with star forming activity,
and non-thermal sources, that are produced by the accretion of
matter onto a Super-Massive Black Hole (SMBH) in the nucleus of the
galaxies--the so called Active Galactic Nuclei (AGNs).

The majority of the emission line galaxies in the SDSS turn out to
be star forming galaxies (SFGs). However, recent studies (e.g.
Miller et al. 2003, Martinez et al. 2010 and Torres-Papaqui et al.
2011) inferred that the actual number of galaxies with a non-thermal
ionization source in its nucleus could be much higher than
previously believed, but the evidence is somewhat obscured by the
large variation of characteristics presented by AGNs. For example,
in the Seyfert~1, where we can distinguish broad emission line
components akin to what is observed in quasars
\citep{osterbrock89,weedman86,krolik99}, the presence of a SMBH
seems indubitable. However, a significantly higher fraction of AGNs
shows only narrow emission lines. Also, many researchers have
alluded to different characteristics for these Narrow Line AGNs
(NLAGNs) by separating them into two main groups, the high
ionization galaxies, generally called Seyfert~2 (Sy2), and the low
ionization ones, called LINERs, which stands for Low Ionization
Nuclear Emission-line Regions \citep{heckman80,coziol96,kewley06}.
As a consequence, the source of ionization of NLAGNs, although
clearly distinct from SFGs, is still an actively debated subject,
and there is presently no consensus about their nature or
evolutionary status.

It is usually admitted that both Sy2 and LINERs are phenomena that
occur in early-type (Sa and Sb) spiral galaxies in the field, that
is, galaxies forming in relatively low galactic density environments
\citep{heckman80}. On the other hand, there is growing evidence that
NLAGNs in clusters and compact groups of galaxies differ from those
in the field by their earlier morphological types and intrinsic low
luminosities
\citep{phillips86,Coz98a,miller03,wake04,martinez08,martinez10,TP11}.
These differences seem to point to distinct formation mechanisms for
galaxies in different environments, which could have also affected
the formation and evolution of their black holes. For example, the
Low Luminosity AGNs (LLAGNs) found in compact groups and clusters of
galaxies appear to prefer massive bulge spiral galaxies or
elliptical galaxies--galaxies which are poor in gas--and it is this
actual scarcity of matter that can be accreted that is assumed to
explain the low luminosity of these objects \citep{martinez08,TP11}.
This explanation is fully consistent with the starving quasar model
\citep{richstone98,gavignaud08,martinez08,martinez10}, according to
which a SMBH that had accreted at high rates in the past, producing
a quasar-like activity, evolved as its reservoir of gas is depleted
into a slowly accreting SMBH producing a LLAGN.

In order to test the idea of a variation of the AGN phenomenon in
different environments we considered important to study galaxies for
which the influence of their environment was minimal. These are
galaxies that are relatively isolated--that is, they formed in low
galactic density environments and evolved without major interactions
with other galaxies.

\section{Description of the sample and data used for analysis}
\label{sec:sample}

Our sample of isolated galaxies was selected from the 2 Micron
Isolated Galaxies catalogue (hereafter 2MIG), as compiled by
\citet{karachentseva10}. The 2MIG catalogue contains 3227 galaxies,
covering the entire sky. All the galaxies have a near-infrared
magnitude brighter than $K_{s}=12$, and a projected image that
extends across an angular diameter $a_{K}\ge 30''$. The isolation
criterion used in this catalogue is the following: a galaxy is
considered isolated when its nearest neighbor has a size within a
factor 4 of the major-axis diameter of the target galaxy, and lies
more than 20 diameters away from it. This criterion assures that a
galaxy with a typical diameter of 20 kpc and peculiar velocity of
the order of {150 km s$^{-1}$} was not influenced by a similar type
of galaxy during the last $\sim 3$ Gyr \citep{turner79}.

To produce our sample, we have cross-correlated the positions of the
galaxies in the 2MIG catalogue with spectroscopic targets in the
Sloan Digital Sky Survey Data Release 7 (SDSS DR7) database
\citep{abazajian09}. This resulted in 445 galaxies, which represents
88\% of the galaxies in the 2MIG catalogue covered by the SDSS DR7.
The remaining 12\% are galaxies that were not observed by SDSS in
spectroscopy, because they were too nearby or too bright.

Color images of the galaxies were obtained using the tool
\begin{scriptsize}CHART\end{scriptsize} of
SDSS\footnote{http://cas.sdss.org/dr7/en/tools/chart/chart.asp}.
After visual inspection, we eliminated 28 galaxies with
morphological peculiarities or possibly faint companions, in
apparent contradiction with the isolation assumption.

A preliminary examination of the spectra reveals that all these
galaxies except one show emission lines. The only non-emission
``isolated'' galaxy is identified as SDSS J072635.39+431746.8. This
is a nearby galaxy, $z=0.0105$, with a possible E or S0 morphology.
Being the only galaxy of its kind it was excluded from our study.

Also excluded from our analysis are two extremely blue S0-like
galaxies, with spectra typical of late-type spirals--showing a
recent and extremely intense level of star formation. A careful
examination of their images suggests they rather are low mass
irregular or morphologically peculiar galaxies. Their positions in
the standard diagnostic diagram reveal they are HII galaxies--small
mass and low gas metallicity starburst galaxies (see Coziol 1996).

\begin{figure*}[!t]
\includegraphics[width=0.95\textwidth]{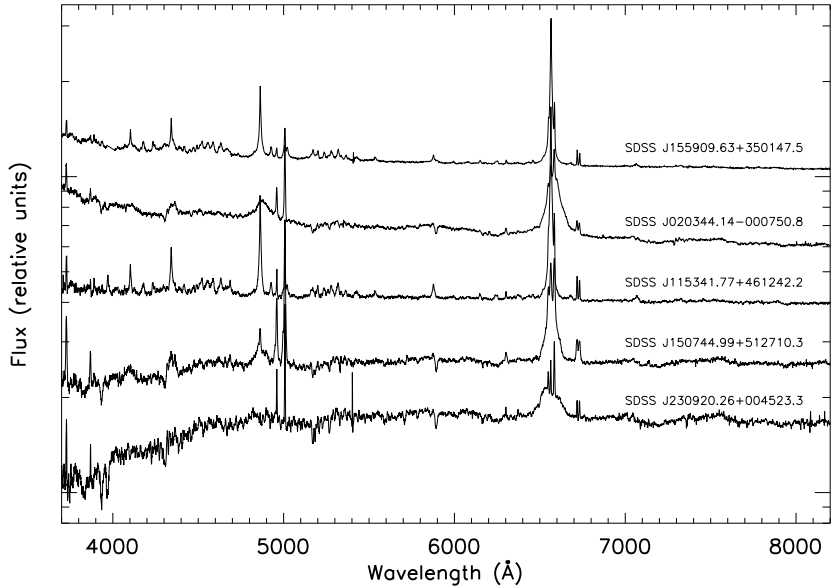}
\caption{Spectra of the broad line AGNs in our original sample of
2MIG galaxies. The flux scale is relative and was slightly displaced
for clarity.} \label{broadem}
\end{figure*}

Only five galaxies in our sample were found to have broad emission
line components. The spectra of these galaxies are presented in
Figure~\ref{broadem}. These galaxies were also discarded from our
analysis.

We classified the 409 remaining galaxies as isolated Narrow Emission
Lines Galaxies (NELGs). The spectra of these galaxies were
subsequently corrected for Galactic extinction \citep{schlegel98},
shifted to their rest frame, resampled to $\Delta \lambda$ = 1\AA\
between 3400 and 8900\AA, and processed using the spectral synthesis
code \begin{scriptsize}STARLIGHT\end{scriptsize} \citep{cid05},
which produces a stellar population template that fits the continuum
of each galaxy. We have run
\begin{scriptsize}STARLIGHT\end{scriptsize} using a combination of
N$_\star$ Simple Stellar Populations (SSPs) from the evolutionary
synthesis models of \citet{bruzual03}. The models were computed with
the
\begin{scriptsize}MILES \end{scriptsize}library \citep{vazdekis10},
following Padova's evolutionary tracks with the initial mass
function of \citet{chabrier03}. An SSP consists of N$_\star = 150$
elements, spanning six metallicities, $Z = 0.005, 0.02, 0.2, 0.4,
1$\ and $2.5 Z _{\sun}$, and 25 stellar population ages in the range
from 1 Myr to 18 Gyr.

In the template-subtracted spectra, we measured automatically
different important attributes of the spectral emission lines, like
their fluxes and Full Width at Half Maximum (FWHM). Other important
features retrieved from the stellar population templates fitted by
\begin{scriptsize}STARLIGHT \end{scriptsize} are the stellar
velocity dispersions and the star formation history (SFH) of the
host galaxies. The SFH map shows how the smoothed Star Formation
Rate (SFR) in a galaxy varies over a time period covering $\log(t) =
5.7$  yrs to $\log(t) = 10.6$ yrs \citep{asari07}.

\subsection{Morphologies and star formation history}
\label{sec:morphologySFH}

\begin{figure*}[!t]
\includegraphics[width=0.95\textwidth]{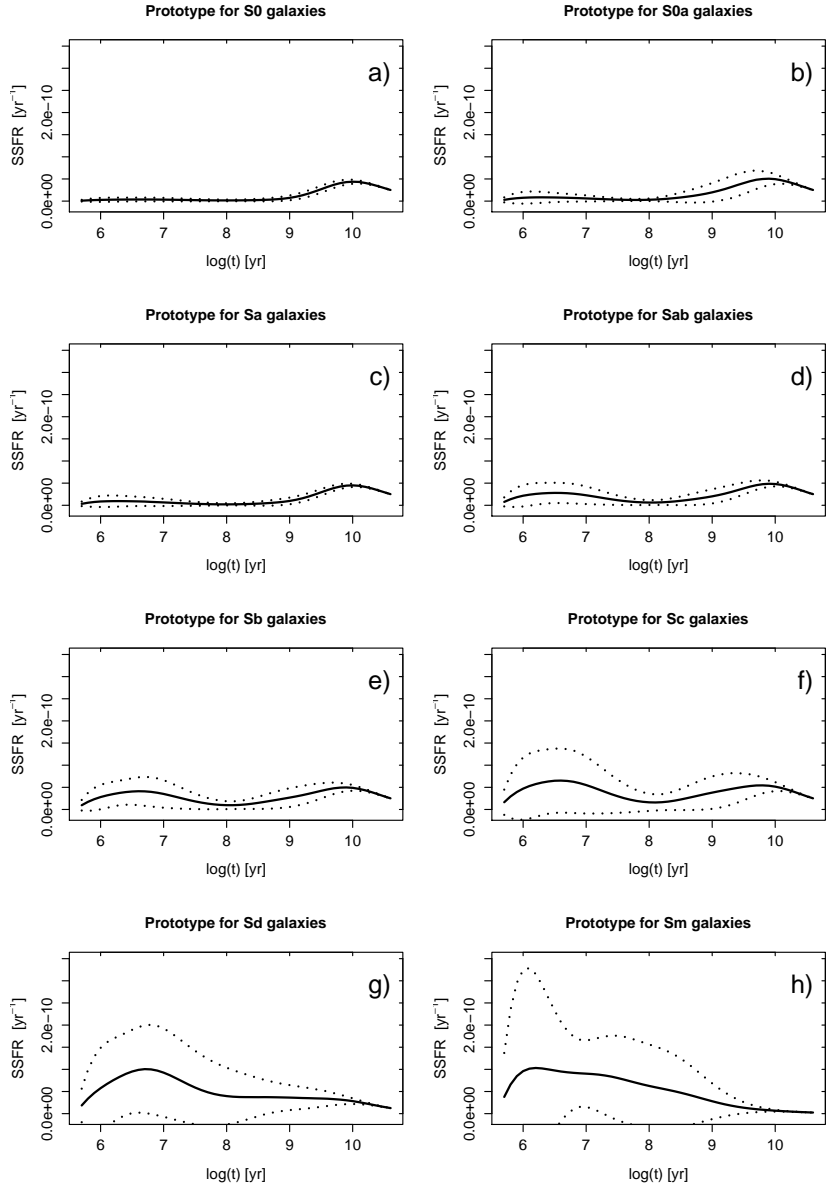}
\caption{SFH for the isolated NELGs. In each graph we show the mean
prototype (solid line) and its dispersion at two sigma (dotted
line). In order to use the same scale for all morphological classes,
we have applied a factor of 0.5 and 0.1 to the SFH curves of the Sd
and Sm, respectively.} \label{ProtosClean}
\end{figure*}

To determine the morphologies of the galaxies we used a composite
method combining an eye estimate with the SFH. First, two of us
(IP-F and RC) determined by eye the morphologies of the 409 isolated
NELGs following the standard Hubble classification. Then, using only
the galaxies with concordant morphologies--meaning the two observers
gave exactly the same morphology--an average SFH prototype for each
morphology class was created. Using these prototypes, the residuals
of the SFH for each galaxy were calculated separately in three
distinct time periods: Recent, $5.7\le \log(t/yrs) \le 7.9$,
Intermediate, $7.9 < \log(t/yrs) \le 9.0$, and Old, $9.0 <
\log(t/yrs) \le 10.6$. A correction was then made to the eye
morphology classification by choosing a morphology that minimizes
the variance of the residuals in each time period. For example, by
eye we could have given the type S0 or Sa to a galaxy, but the
residuals being minimal for Sa we adopted this last classification.

Once the morphologies were correctly adjusted to the SFH, we
recalculated the prototypes by calculating the mean SFH in each
morphological bin. Because we found the Sb and Sbc types to have SFH
prototypes that are extremely similar and almost impossible to
distinguish, we merged these two morphological bins together,
keeping only the Sb identification. The final prototypes, including
327 (80\%) of the 409 isolated NELGs, are presented in
Figure~\ref{ProtosClean}. The prototypes suggests S0, S0a, and Sa
galaxies formed most of their stars in the past, $\log(t) > 9.0$
yrs, while the Sab had near constant star formation activity over
the whole time lapse covered by \begin{scriptsize}STARLIGHT
\end{scriptsize}. This is also true for the Sb, but with a slightly
larger dispersion. In the Sc, recent star formation activity becomes
more important than in the past, but the dispersion also increases
significantly with earlier types. In Sd and Sm galaxies most of the
stars appear to have formed recently, $\log(t) < 8.0$  yrs. These
results are in good agreement with similar studies based on
integrated spectroscopy (e.g. Kennicutt 1992).

The remaining 20\% galaxies were considered peculiar. Among these
galaxies we count 36 late-type spirals (Sc or Sd) with a SFH typical
of an Sa. These galaxies turned out to be cases where the fiber was
covering only the innermost part of the galaxies. Also considered
peculiar are 22 galaxies with a burst of star formation $\sim 100$
Myr in the past. No external cause for the bursts could be
determined, the galaxies being confirmed as isolated. Another 24 E
or S0 galaxies were found to show a trace (not intense) of very
recent star formation. All these peculiar galaxies were excluded
from our analysis.

\subsection{Activity classification}

\begin{table*}[!t]\centering
  \small
  \begin{changemargin}{-1.5cm}{-1.5cm}
  \caption
{Examples of spectroscopic characteristics of the isolated NELGs.
\lowercase{(\textsc{C}omplete table available in electronic form.)}}
  \begin{tabular}{lcccccc}
  \hline
Identification in SDSS   & $\log\left(\frac{\textrm{O[III]}}{\textrm{H$\beta$}}\right)$ & $\log\left(\frac{\textrm{N[II]}}{\textrm{H$\alpha$}}\right)$ & Activity & L$_{H\alpha}$ & $\sigma_\star$ & $\log(\lambda$L$_{5100})$\\
                         &                                                              &                                                              &  Type    & L$_\odot$     &  km s$^{-1}$   & erg s$^{-1}$ \\
\hline
SDSS J105809.84$-$004628.8 &  $-0.364$ &  $-0.351$ &  SFG  & 6.19 &    87   & 42.13    \\
SDSS J113903.33$-$001221.6 &  \ \ 0.228  &    \ \ 0.060  &  AGN  & 6.05 &    144  & 42.56    \\
SDSS J135807.05$-$002332.9 &  $-0.604$ &  $-0.521$ &  SFG  & 7.03 &    86   & 42.76    \\
SDSS J142223.76$-$002315.5 &  $-0.437$ &  $-0.494$ &  SFG  & 5.50 &    81   & 41.16    \\
SDSS J150654.85$+$001110.8 &    \ \ 0.253  &    \ \ 0.055  &  AGN  & 6.67 &    202  & 43.11    \\
SDSS J113423.32$-$023145.5 &    \ \ 0.363  &    \ \ 0.238  &  AGN  & 6.23 &    136  & 43.00    \\
SDSS J115425.04$-$021910.3 &  $-0.275$ &  $-0.453$ &  SFG  & 5.36 &    78   & 41.29    \\
SDSS J122353.98$-$032634.4 &  $-0.218$ &  $-0.148$ &  TO   & 5.39 &    113  & 41.64    \\
SDSS J124428.12$-$030018.8 &  $-0.193$ &  $-0.300$ &  TO   & 6.28 &    115  & 42.69    \\
SDSS J170128.21$+$634128.0 &  $-0.671$ &  $-0.375$ &  SFG  & 7.20 &    99   & 42.52    \\
\hline \label{TabBPT}
\end{tabular}
\end{changemargin}
\end{table*}

For our spectral activity analysis we further reduce our sample to
292 NELGs that have a S/N\ $\ge 3$\ in the four most important
emission lines: H$\beta$, [OIII]$\lambda$5007, H$\alpha$ and
[NII]$\lambda6584$. In Table~\ref{TabBPT} we compile the
spectroscopic characteristics of these galaxies as measured in the
template subtracted spectra. Column~1 gives the SDSS identification
of the galaxies. In columns~2 and 3 we put the values of the two
line ratios used to determine the activity type of the galaxies. The
adopted classification is given in column~4. In column~5 we list the
H$\alpha$ emission line luminosity. In column~6 we give the velocity
dispersion of the stars, $\sigma_\star$, as deduced from the
\begin{scriptsize}STARLIGHT \end{scriptsize}template and corrected
for the SDSS spectral resolution. Considering the size of the fiber
used in SDSS (3"), this value can be taken as the stellar velocity
dispersion of the bulge of the galaxy. Finally, in column~7, we give
the luminosity of the continuum at $\lambda$5100\ \AA\ as measured
in the spectra before template subtraction.

\begin{figure*}[!t]
\includegraphics[width=0.95\textwidth]{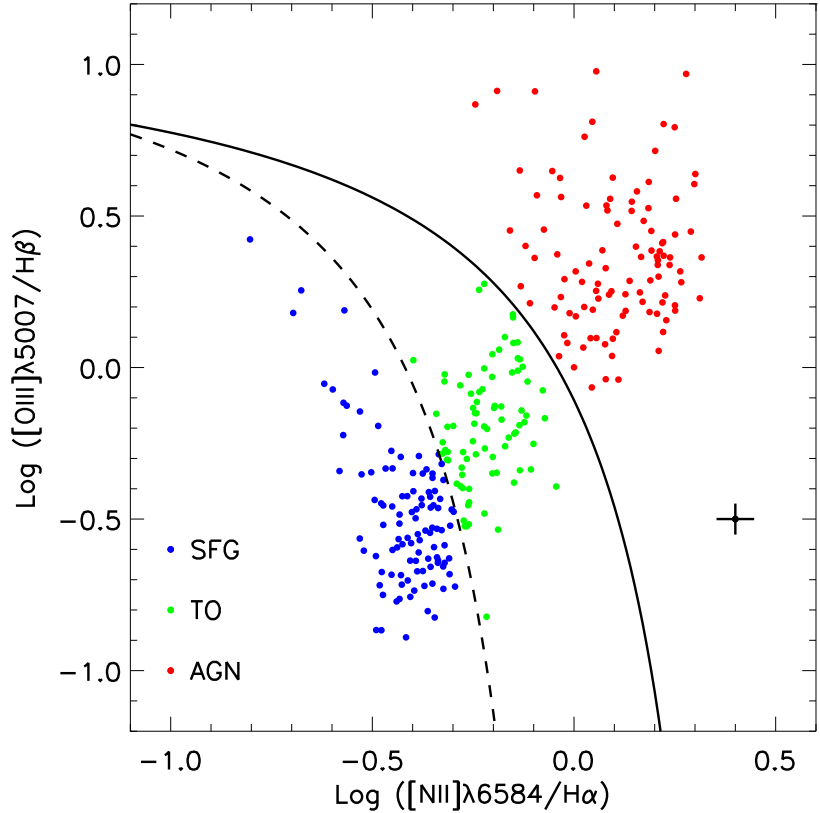}
\caption{Diagnostic diagram for the activity classification of the
isolated NELGs. The separation between SFGs and TOs was suggested by
\citep{kauffmann03} and the separation between TOs and AGNs was
suggested by \citep{kewley06}. Mean uncertainties on the line ratios
(not a data) are indicated as a cross on lower right of diagram.}
\label{bptdiagram}
\end{figure*}

We have identified the activity type of the 292 isolated NELGs in
Figure~\ref{bptdiagram}, where we compare the two line ratios
[OIII]$\lambda$5007/H$\beta$ and [NII]$\lambda6584$/H$\alpha$. The
typical low uncertainty levels on these ratios (cross in
Figure~\ref{bptdiagram}) have practically no effect on our
classification. We distinguish 105 SFG galaxies (36.0\%), 83 TOs
(28.4\%), and 104 AGNs (35.6\%).

\subsection{Physical characteristics of the isolated NELGS}

\begin{table*}[!t]\centering
  \small
  \begin{changemargin}{-2.0cm}{-2.0cm}
\caption{Examples of basic physical properties for the isolated NELGs.\\
\lowercase{(\textsc{C}omplete table available in electronic form.)}}
\begin{tabular}{lccccccccccc}
\hline
Identification in SDSS & RA (J2000)& DEC (J2000)&  z        & Morph.&  $R_{90\%}$ & CI & $\log$(L$_K$)& Mean $(t_*)$  \\
                & (degree)  & (degree)   &           &       &  Kpc        &    & L$_\odot$    &   yrs         \\
\hline
SDSS J105809.84$-$004628.8 &  164.54102 &  $-0.77468$ &    0.0215 &  Sc   &    6.5 &  2.18 &  10.43 &  8.79     \\
SDSS J113903.33$-$001221.6 &  174.76388 &  $-0.20600$ &    0.0181 &  Sab  &    8.6 &  3.06 &  10.77 &  9.82     \\
SDSS J135807.05$-$002332.9 &  209.52940 &  $-0.39248$ &    0.0296 &  Sc   &    9.8 &  2.11 &  10.80 &  8.61     \\
SDSS J142223.76$-$002315.5 &  215.59903 &  $-0.38766$ &    0.0055 &  Sd   &    4.6 &  1.90 &  10.01 &  7.81     \\
SDSS J150654.85$+$001110.8 &  226.72857 &    \ \ 0.18634  &    0.0351 &  Sa   &   12.1 &  2.72 &  11.19 &  9.60     \\
SDSS J113423.32$-$023145.5 &  173.59719 &  $-2.52933$ &    0.0395 &  Sb   &   13.9 &  2.09 &  11.06 &  9.79     \\
SDSS J115425.04$-$021910.3 &  178.60434 &  $-2.31955$ &    0.0080 &  Sd   &    4.5 &  2.12 &   9.78 &  8.05     \\
SDSS J122353.98$-$032634.4 &  185.97495 &  $-3.44292$ &    0.0068 &  Sab  &    5.5 &  2.79 &  10.58 &  8.92     \\
SDSS J124428.12$-$030018.8 &  191.11718 &  $-3.00525$ &    0.0239 &  Sc   &    8.6 &  2.65 &  10.61 &  9.17     \\
SDSS J170128.21$+$634128.0 &  255.36758 &    \ 63.69112 &    0.0163 &  Sc   &    5.5 &  1.76 &  10.42 &  7.89     \\
\hline \label{Tabbasic}
\end{tabular}
\end{changemargin}
\end{table*}

We present the basic physical properties of the 292 isolated NELGs
in Table~\ref{Tabbasic}. After the SDSS identification in column~1,
we list in columns~2, 3 and 4 the right ascension, declination and
redshift of the galaxies as listed in SDSS DR7. The results of our
morphological classification are given in column~5. Also from the
SDSS DR7 database, we give in columns~6 and 7 the Petrosian radii
(in kpc) at 90\% of the light distribution and the concentration
index, CI, which is the ratio of the two Petrosian radii (CI\ $ =
R_{90\%}/R_{50\%}$). In column~8 we put the luminosity in the K band
as determined from 2MASS (K20 from 2MASS). In the last column,
column~9, we give the mean stellar age of the stellar populations
consistent with the
\begin{scriptsize}STARLIGHT \end{scriptsize}template
\citep{asari07}.

The K20 magnitudes from 2MASS were extracted from the extended
sources catalog \citep{jarrett00} by choosing the closest galaxy
within 5 arcsecs of the best match. They were corrected for Galactic
extinction at the position of the galaxies using the dust maps
published in \citet{schlegel98}. We have also applied a {\it
k}-correction using the method described in \citet{kochanek01}. This
method is reported to be valid for $ z <0$.25 and to be independent
of the galaxy morphological type. The K band absolute magnitude was
calculated using the standard relation:
\begin{equation}
 M_K  = m_K - 25 - 5 \log(D_L) - A_K - k(z) ,
\end{equation}
where $m_K$ is the total apparent magnitude in K, $A_K$ is the
Galactic extinction, $D_L$  is the luminosity distance in Mpc,
calculated assuming H$_{0} = 75$ km s$^{-1}$ Mpc$^{-1}$, and $ k (z)
=-6.0\log(1+z)$ is the {\it k}-correction. From the magnitudes we
deduce the K luminosities applying the relation:
\begin{equation}
 \log (L_K/L_\odot)  = 0.4(M_{\sun K} - M_K) ,
\end{equation}
where M$_{\sun K}=3.28$\ is the absolute magnitude of the Sun in the
K band \citep{binney98}. From the uncertainties on the fluxes
\citep{jarrett00}, a mean uncertainty of 6\% is estimated for our K
luminosities.

\begin{figure*}[!t]
\includegraphics[width=0.95\textwidth]{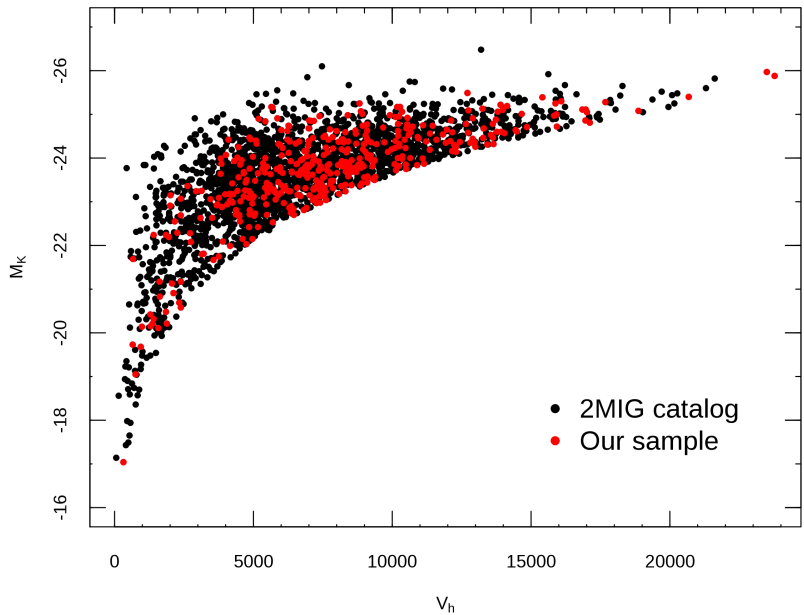}
\caption{Comparison of absolute magnitude in K vs. heliocentric
velocity in the 2MIG catalog and our subsample of 292 isolated
NELGs. } \label{velK}
\end{figure*}

In Figure~\ref{velK} we compare the distributions of the 2MASS
absolute magnitudes in K and heliocentric radial velocities, as
found in our subsample of 292 and the entire 2MIG catalogue. The
distribution for the 2MIG catalogue is typical of a flux limited
survey. The galaxies in our subsample are randomly distributed,
suggesting they form a statistically fair sample of all the isolated
galaxies in the 2MIG catalogue.

\section{Analysis}

\subsection{Activity type vs morphology}

\begin{figure*}[!t]
\includegraphics[width=0.95\textwidth]{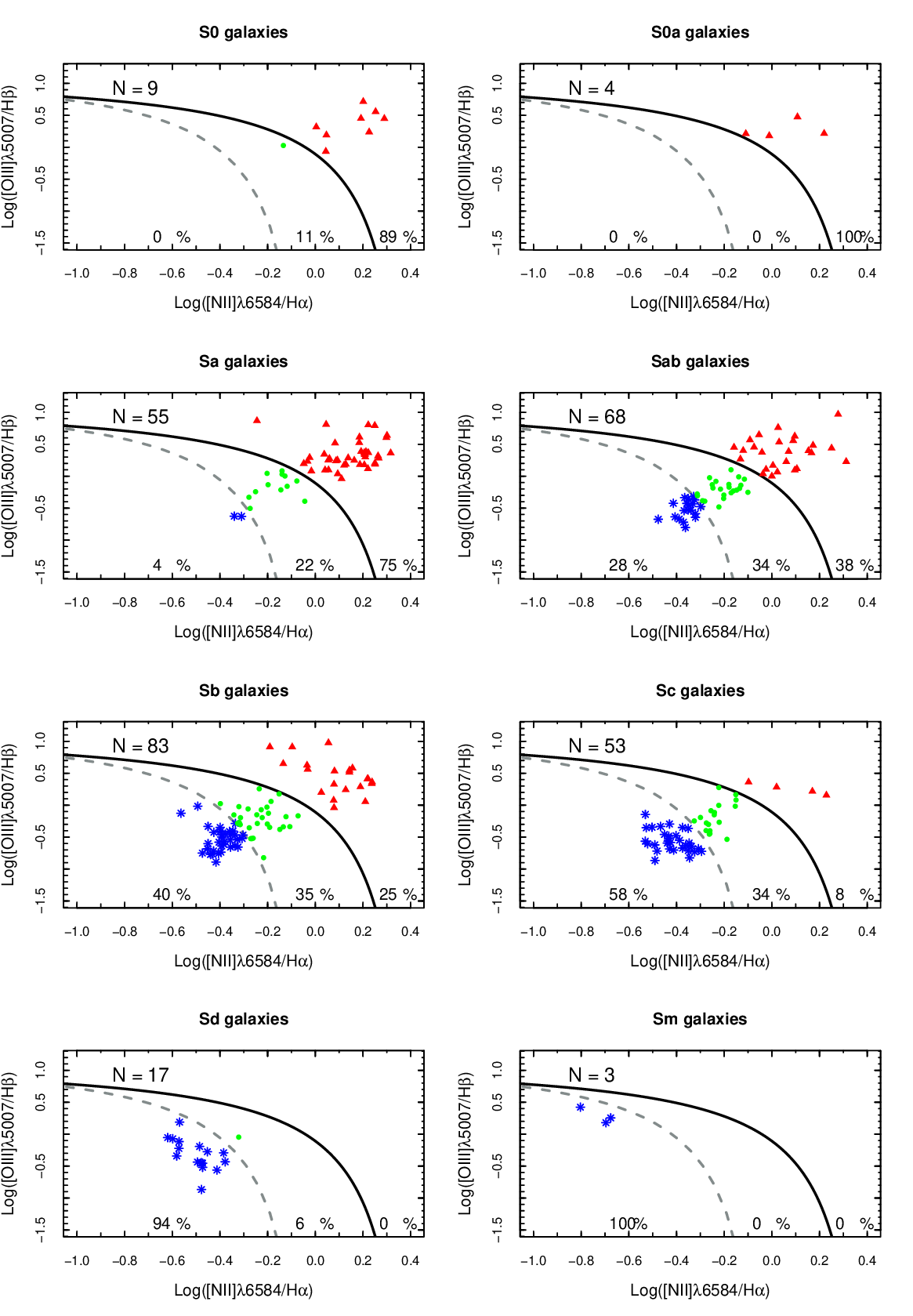}
\caption{Diagnostic diagrams for isolated NELGs with different
morphologies. The separations and colors are as explained in
Figure~\ref{bptdiagram}.} \label{actmorph}
\end{figure*}

We present the fractions of morphology types in our sample of 292
isolated NELGs in Table~\ref{TabMorphFraction}. The majority
(70.5\%) of these galaxies are classified as intermediate spirals
(Sa-Sab-Sb). In Figure~\ref{actmorph} we show how the distribution
of activity types varies in the different morphology classes: AGNs
and TOs are mostly found in early-type galaxies (S0-Sa), while the
proportion of SFGs gradually increases in later types (Sab-Sm). This
result is not new, but confirms the common view about AGNs in the
field: they are mostly found in early-type spiral galaxies (e.g.
Melnick, Terlevich \& Moles 1986; Osterbrock 1989; Blandford, Netzer
\& Woltjer 1990).

\begin{table*}[!t]\centering
  \caption{Distribution of morphologies in 292 isolated NELGs.}
  \begin{tabular}{ccc}
  \hline
   Morphology  &  Number          &  Fraction          \\
      type     &                  &    \%              \\
\hline
S0             &          9       &         3.1        \\
S0a            &          4       &         1.4        \\
Sa             &          55      &         18.8       \\
Sab            &          68      &         23.3       \\
Sb             &          83      &         28.4       \\
Sc             &          53      &         18.2       \\
Sd             &          17      &         5.8        \\
Sm             &          3       &         1.0        \\

\hline

 \label{TabMorphFraction}
\end{tabular}
\end{table*}

\begin{table*}[!t]\centering
  \caption{
Examples of masses for the isolated NELGs. \\
\lowercase{(\textsc{C}omplete table available in electronic form.)}}
  \begin{tabular}{lccc}
  \hline
Identification in SDSS &  \multicolumn{3}{c}{$\log$(\textit{M}) (M$_\odot$)}           \\
                       &  \textit{M}$_{Bulge}$          & \textit{M}$_{BH}$         &  \textit{M}$_{B}$  \\
\hline
SDSS J105809.84$-$004628.8  &   9.04    &           & 10.71 \\
SDSS J113903.33$-$001221.6  &   9.40    &   6.41    & 10.55 \\
SDSS J135807.05$-$002332.9  &   9.17    &           & 11.04 \\
SDSS J142223.76$-$002315.5  &   8.39    &           & 10.17 \\
SDSS J150654.85$+$001110.8  &   9.99    &   7.06    & 11.12 \\
SDSS J113423.32$-$023145.5  &   9.69    &   6.74    & 11.22 \\
SDSS J115425.04$-$021910.3  &   8.52    &           &  9.86 \\
SDSS J122353.98$-$032634.4  &   8.77    &   5.70    & 10.28 \\
SDSS J124428.12$-$030018.8  &   9.33    &   6.33    & 10.96 \\
SDSS J170128.21$+$634128.0  &   9.03    &           & 10.77 \\
\hline \label{MASS}
\end{tabular}
\end{table*}

\begin{figure*}[!t]
\includegraphics[width=0.95\textwidth]{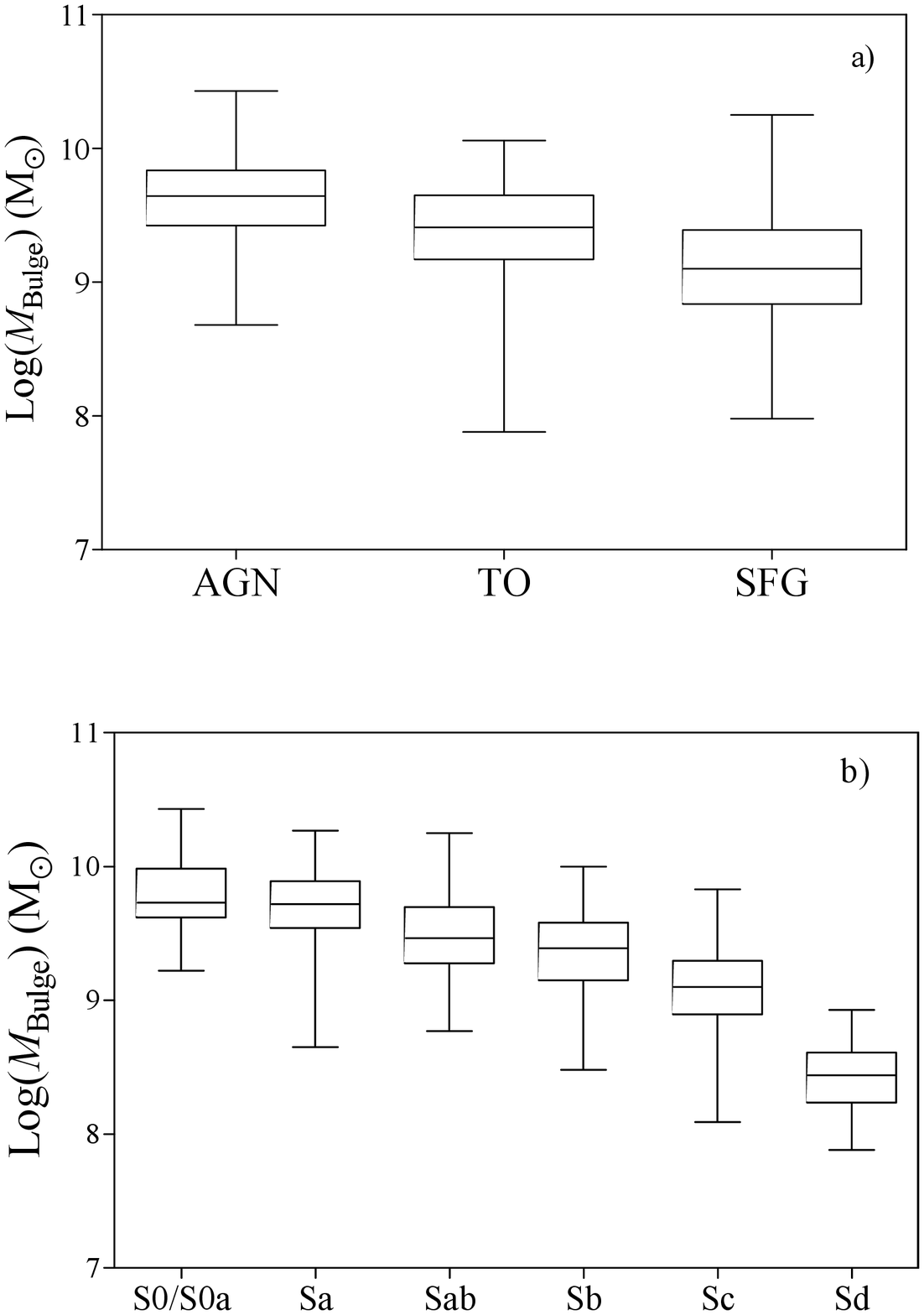}
\caption{Box-whisker plots for the bulge masses of galaxies as
function of a) activity type, b) morphology. The middle line is the
median, the box is limited by the percentiles and the whiskers
indicate the full range of values.} \label{BWbulges}
\end{figure*}

In Table~\ref{MASS} we give our estimated masses for the bulges,
$M_{Bulge}$, as determined using the velocity dispersions,
$\sigma_\star$, and applying the virial theorem. The box-whisker
plots for the bulge masses of the galaxies separated by activity
types is presented in Figure~\ref{BWbulges}a. The median values are
reported in Table~\ref{Median_Mass}. AGNs and TOs possess more
massive bulges than SFGs. The trends observed for the AGNs and TOs
are confirmed using a non-parametric statistical test
(Kruskall-Wallis with Dunn's multiple comparison tests). The results
of these tests are presented in Table~\ref{test1} of the appendix.
The tests find the medians in the three samples to be significantly
different at a level of confidence of 99\%.

In Figure~\ref{BWbulges}b we compare the bulge mass of galaxies with
different morphologies. We also observe a significant difference,
the bulge mass being higher in earlier morphological types. This
trend is also found to be statistically significant. The results for
the tests, presented in Table~\ref{test2} of the appendix, show that
the farther apart in morphological class, the more significant the
difference in bulge mass. For example, the statistical tests reveal
no difference between the Sa and Sab, but a significant one between
the Sa and Sb. Similarly, no difference appears between the Sab and
Sb, but a significant one is found between the Sab and Sc.

\begin{figure*}[!t]
\includegraphics[width=0.95\textwidth]{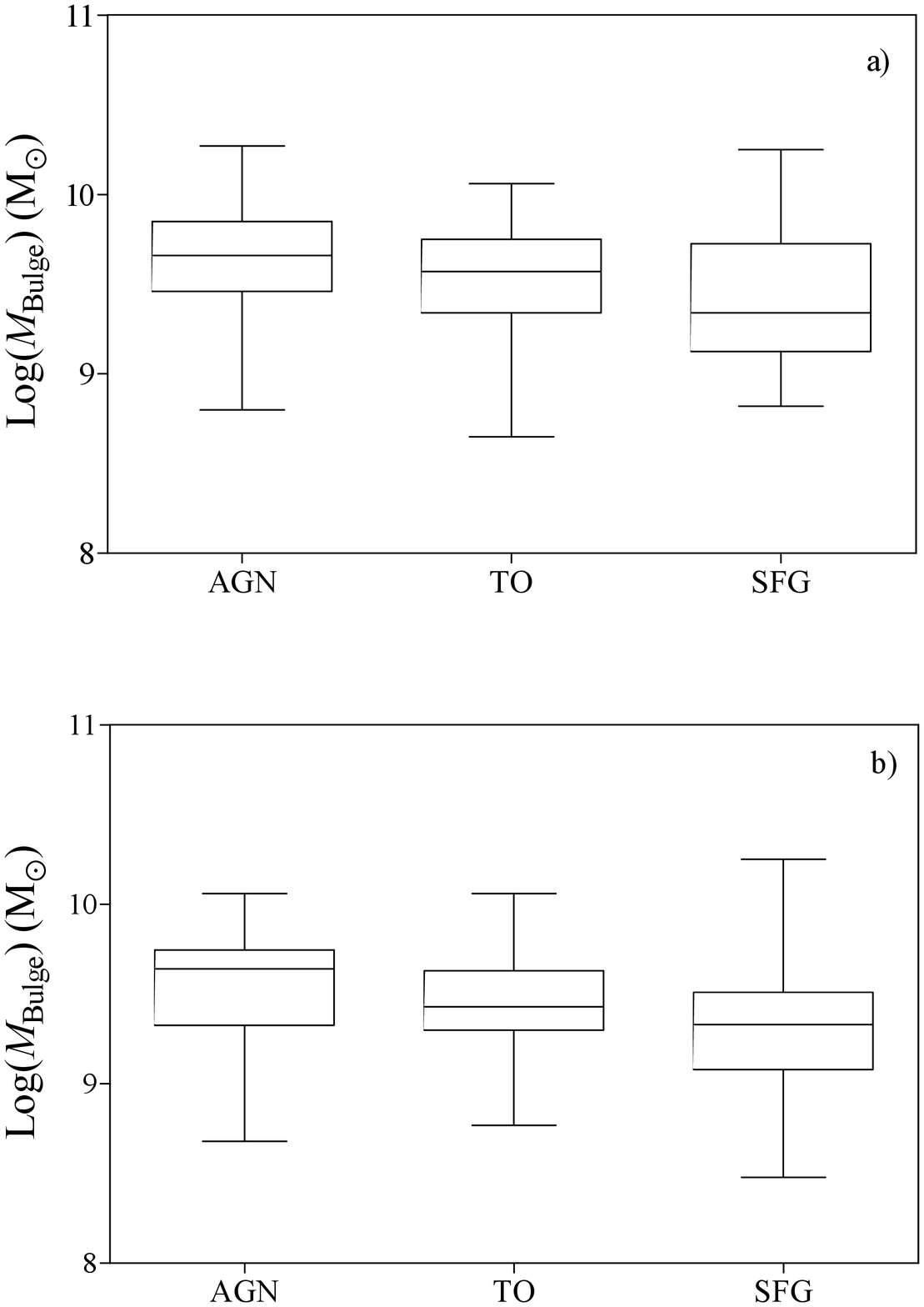}
\caption{Box-whisker plots for the bulge masses as found a) in group
1 (Sa and Sab galaxies), and b) in group 2 (Sab and Sb galaxies).
Box-whiskers are defined as in Figure~\ref{BWbulges}.}
\label{BWbulgesvsM}
\end{figure*}

The above results lead to the question of whether the most important
parameter is the mass of the bulge or the morphology of the galaxy.
As a test, we compare the bulge masses in two statistical groups:
group~1 which is composed of Sa and Sab galaxies, and group~2 which
is composed of Sab and Sb. The box-whisker plots are drawn in
Figure~\ref{BWbulgesvsM}a for group~1 and Figure~\ref{BWbulgesvsM}b
for group~2. We find no significant difference in bulge mass between
the different activity types in group~1, while in group 2 there is a
significant difference only between the SFGs and AGNs. This is
confirmed by the statistical tests in Table~\ref{test1} of the
appendix. We conclude that the AGN activity is related to the bulge
mass, but mostly because of the strong connection of this parameter
with the morphology: the earlier the morphology, the bigger the
bulge mass, and the higher the probability to see an AGN. This is
also consistent with the diagnostic diagrams, which show a growing
of the frequency of AGNs in earlier morphological types. In group 1,
the numbers of AGNs, SFGs and TOs are 67, 21 and 35, respectively,
while in group 2 these numbers change to 47, 53 and 51 respectively.
For an increase by a factor 2 in bulge mass, from Sa to Sb (the
bulge mass increases by factor 3.5 from AGN to SFG in Table~5), the
number of AGNs slightly falls and the number of TOs and SFGs
doubles.

The bulge mass being an important morphological parameter, the
strong correlation found with the AGN activity suggests this
phenomenon is intimately connected with the formation process of the
galaxies.

\subsection{Activity type vs. galaxy mass}

We have estimated the masses of our galaxies using two different
methods. The first method is based on 2MASS K band luminosities (as
compiled in Table~2), while the second is based on the absolute B
magnitudes, which were obtained using the Johnson-B band magnitude
synthesized from the SDSS magnitudes \citep{fukugita96}. The
absolute magnitudes in B are compiled in Tables~\ref{TableOHSFG},
\ref{TableOHAGN} and \ref{TableOHTO}, for the SFGs, AGNs, and TOs,
respectively. The uncertainty is of the order of 0.05 mag. Being
insensitive to dust extinction and the morphological type of the
galaxies the near-infrared emission is a better tracer of the
stellar mass than the B magnitudes, for which we need to apply a
correction depending on the morphology.

\begin{figure*}[!t]
\includegraphics[width=0.95\textwidth]{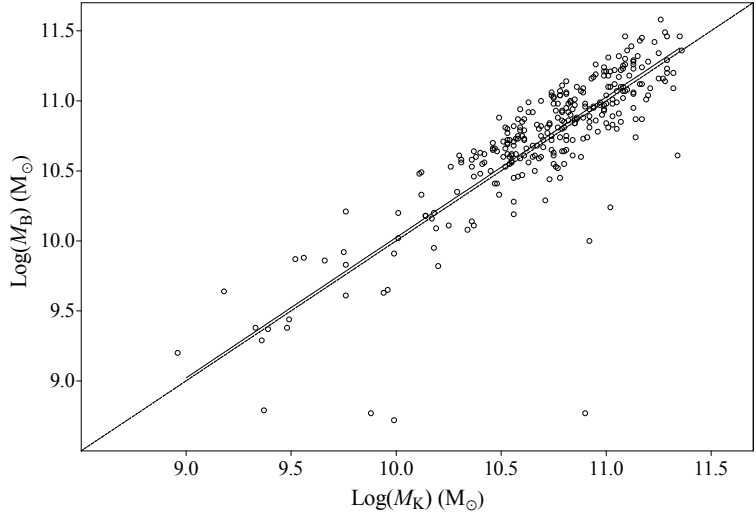}
\caption {Comparison of the two mass estimates. The dot line is the
one to one relation and the continuous line is a linear fit.}
\label{MASScomparison}
\end{figure*}

The K luminosities were transformed into masses using the
mass-to-light ratio $M/L_K = 0.95$, as estimated by \citet{bell03}.
To transform the B absolute magnitudes into masses, we first
corrected for galactic extinction and {\it k}-correction using the
code developed by \citet{BR07}. Then we applied the different
mass-to-light ratios for galaxies as published by \citet{FG79}.
These ratios were adjusted for our adopted value of the Hubble
constant. The B mass estimates appear in column~4 of
Table~\ref{MASS}.

\begin{table*}[!t]\centering
\caption{Median masses and concentration indices.}
  \begin{tabular}{lcccc}
\hline
          &  \multicolumn{3}{c}{$\log$(\textit{M}) (M$_\odot$)}          & CI \\
          &  \textit{M}$_{Bulge}$          & \textit{M}$_{K}$         &  \textit{M}$_{B}$ &    \\
\hline
AGN       &  9.64 &10.90 & 10.92&2.41 \\
TO        &  9.41 &10.85 & 10.78&2.21 \\
SFG       &  9.10 &10.66 & 10.56&2.11 \\
S0/S0a    &  9.73 &10.98 & 10.94&3.21 \\
Sa        &  9.72 &10.91 & 11.00&2.87 \\
Sab       &  9.46 &10.82 & 10.78&2.62 \\
Sb        &  9.39 &10.94 & 10.77&2.34 \\
Sc        &  9.10 &10.66 & 10.58&2.12 \\
Sd        &  8.44 & 9.86 &  9.66&2.07 \\
\hline
 \label{Median_Mass}
\end{tabular}
\end{table*}

In Figure~\ref{MASScomparison} we compare the two mass estimates.
The linear relation fitted has a correlation parameter $r^2=0.72$,
implying that 72\% of the total variance of the B mass estimates is
explained by the variation of the K mass estimates (and vice versa).
The median values for the masses in the B and K bands are given in
Table~\ref{Median_Mass}. They are in excellent agreement with those
reported by \citet{roberts94} for galaxies having similar
morphological types.

In Table~\ref{Median_Mass} we see that both median masses, $M_B$ and
$M_{K}$, tend to increase from the SFGs to the AGNs. The general
trend observed in the box-whisker plots presented in
Figure~\ref{BW_ATvsMass}, looks slightly more obvious using $M_K$
than $M_{B}$. In Table~\ref{test1} of the appendix we see that using
$M_{B}$ the statistical tests detect a slightly more significant
difference between the AGNs and SFGs than between the TOs and SFGs,
while both differences are similarly statistically significant using
$M_K$. No significant difference is observed between the AGNs and
TOs.

\begin{figure*}[!t]
\includegraphics[width=0.95\textwidth]{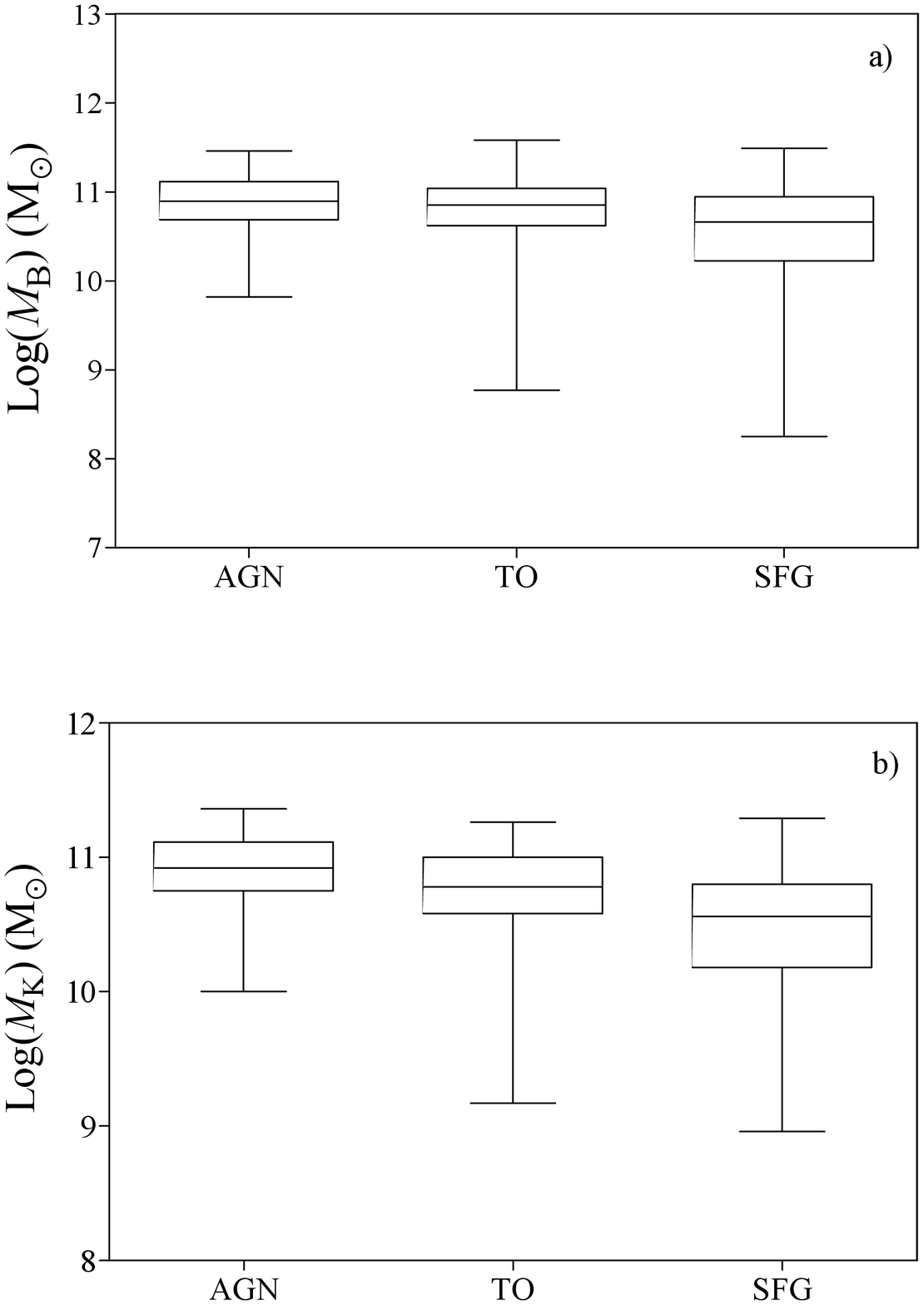}
\caption{Box-whisker plots for the total masses in galaxies showing
different activity types; a) \textit{M}$_{B}$ and b)
\textit{M}$_{K}$.} \label{BW_ATvsMass}
\end{figure*}

\begin{figure*}[!t]
\includegraphics[width=0.95\textwidth]{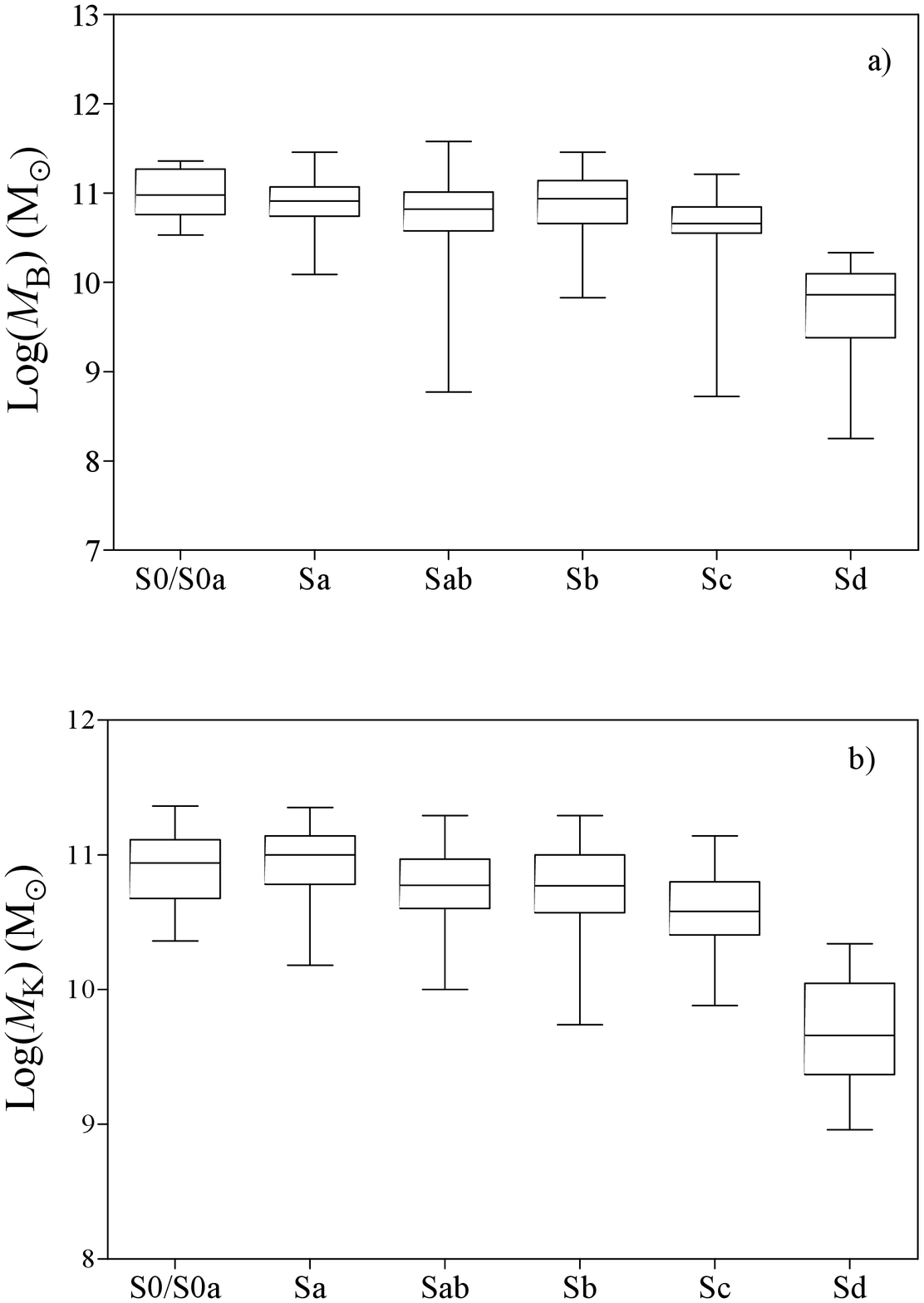}
\caption{Box-whisker plots for the total masses in galaxies having
different morphologies; a) \textit{M}$_{B}$ and b)
\textit{M}$_{K}$.} \label{BW_MvsMass}
\end{figure*}

In Figure~\ref{BW_MvsMass} we compare the masses of the galaxies
having different morphological types. Although we found a
significant variation of the bulge mass with the morphology, the
total mass $M_B$ and $M_K$ do not seem to vary as much between
galaxies having different morphologies. In Table~\ref{test2} of the
appendix, we find that the observed differences start to be
statistically significant only when the comparison is done with the
latest types, Sc for $M_{B}$, and even later, Sd, for $M_K$. We
conclude that the galaxies with different morphologies show only
marginal differences in total masses.

Based on our analysis, the trends observed between the AGN activity,
the bulge mass and morphology does not appear to be ``quantitative''
(based on a difference in total mass), but more ``qualitative'', the
AGN activity appearing more frequently when a higher fraction of the
mass of the galaxy is in the form of a bulge. This seems to favor a
mechanism based on different astration rates--defined as the
efficiency with which a galaxy transforms its gas into stars
\citep{Sandage86}. The higher the astration rate, the more massive
the bulge and the higher the probability to observe an AGN.

\subsection{Relation with stellar populations}

\begin{figure*}[!t]
\includegraphics[width=0.95\textwidth]{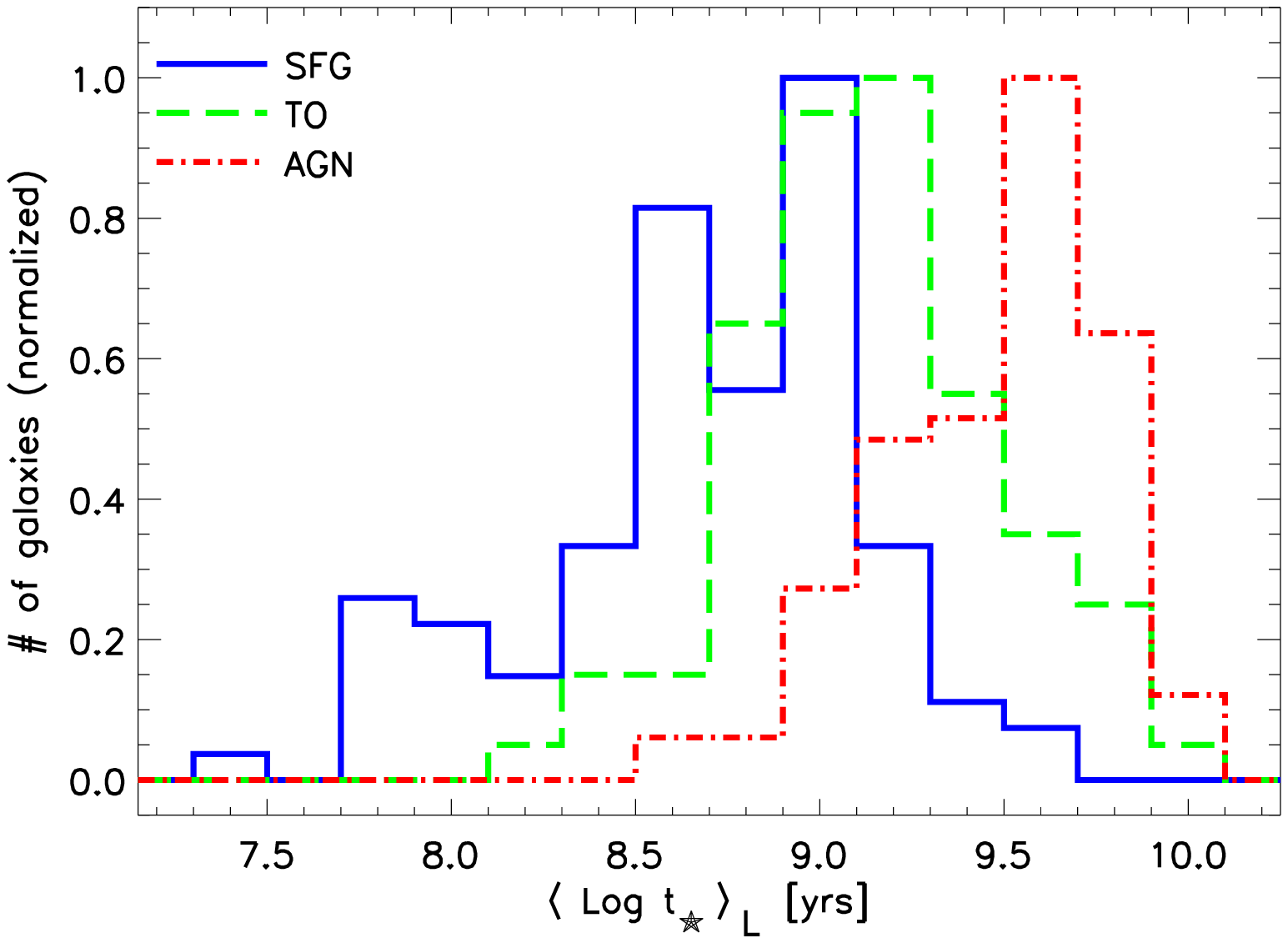}
\caption{Distributions of mean ages of stellar populations in
galaxies with different activity types.} \label{medianage01}
\end{figure*}

\begin{figure*}[!t]
\includegraphics[width=0.95\textwidth]{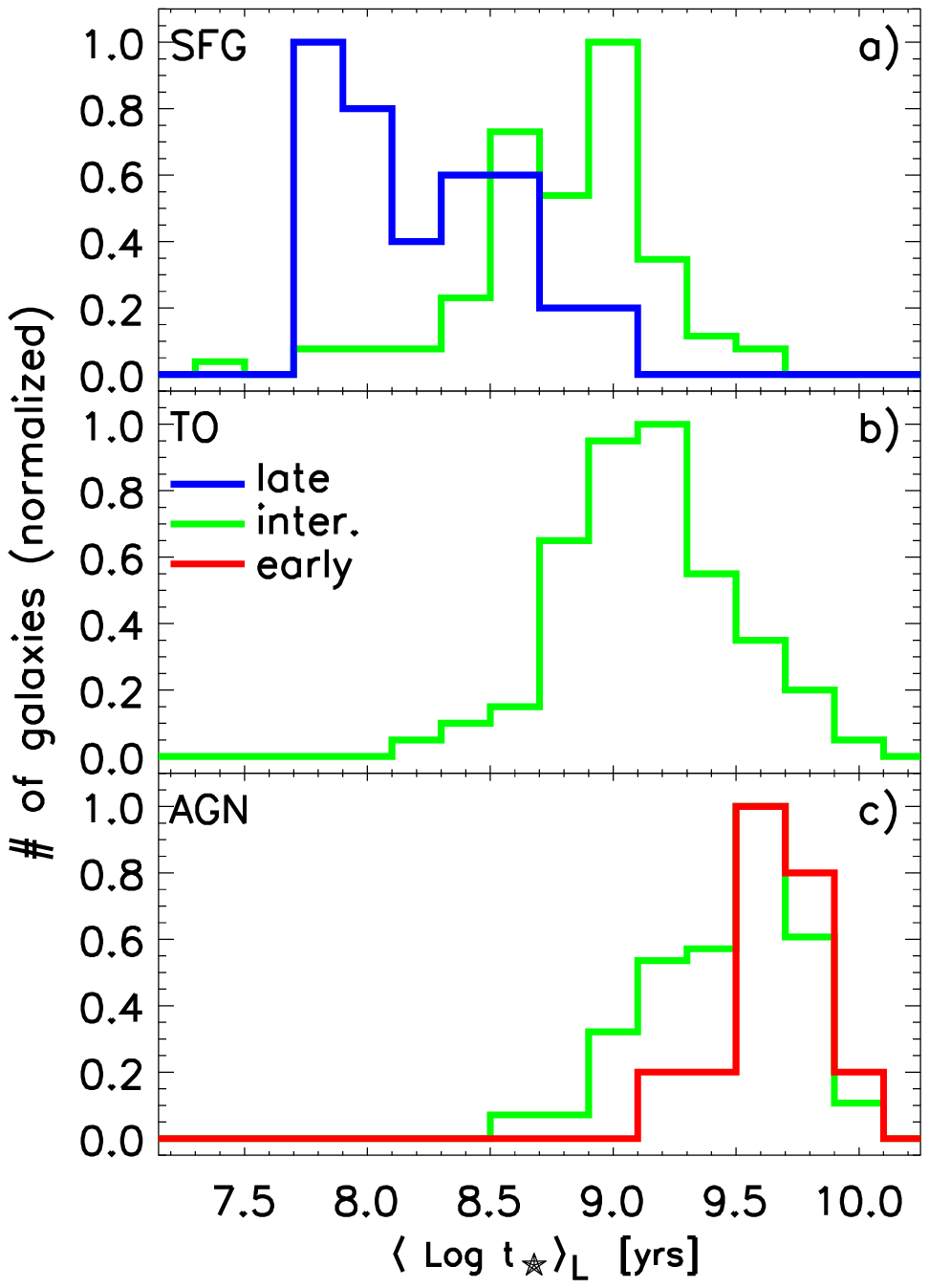}
\caption{Distributions of mean ages of stellar populations in
galaxies with different activity types and different morphologies.}
\label{medianage02}
\end{figure*}

In Figure~\ref{medianage01} we compare the distributions of the mean
ages of the stellar populations in galaxies showing different
activity types. The stellar populations of the SFGs are dominated by
young stars, with mean ages ranging from 0.32 to 1.0 Gyrs. These
values correspond to the 25 and 75 percentiles, respectively. The
mean age of the stellar populations is observed to go up in the TOs,
with values ranging from 0.5 to 3.2 Gyrs, and to culminate in the
AGNs with values ranging from 1.6 to 6.3 Gyrs. Comparable results
were encountered before by \citet{Boisson00}.

A strong connection is also found with the morphologies. In
Figure~\ref{medianage02}, the morphologies were regrouped into three
broad classes: Late (Sc, Sd, and Sm), Intermediate (Sa, Sab, and
Sb), and Early (S0 and S0a). The SFGs show a mixture of Late and
Intermediate morphologies, while the TOs have only Intermediate
morphologies. The AGNs on the other hand have mostly Early
morphologies.

The correlations between morphology, activity type and mean stellar
ages of the stellar populations support that there is a strong
connection between the AGN activity and the formation process of the
galaxies. In particular, the AGN phenomenon appears like a normal
by-product of the formation process of galaxies that produces more
massive bulges.

There are many structural similarities between the bulges of spiral
galaxies and elliptical galaxies, suggesting similar formation
mechanisms (e.g. Jablonka, Martin \& Arimoto 1996). In particular,
elliptical galaxies are known to have endured higher astration rates
than spiral galaxies when they formed \citep{Sandage86},
transforming almost all their gas into stars in a very short period
of time. Assuming that galaxies form by a succession of star forming
episodes--an assumption necessary to produce the mass-metallicity
relation--galaxies with high astration rates would thus be expected
to have formed most of their stars in the past, because the
reservoir of gas being limited, the galaxy would have consumed it
rapidly and stopped forming stars relatively early. This would
explain the differences in mean ages for the stellar populations
observed in our sample of isolated NELGs.

\subsection{Relation with gas metallicity}

After correcting for dust absorption, the metallicities of the gas
in the SFGs was estimated using the empirical correlation
encountered between $\log ($O/H$)$ and the emission-line ratio
R$_3$= ([OIII]$\lambda$4959+[OIII]$\lambda$5007)/H$_\beta$
\citep{EP84,VC92}. The correlation originates from the cooling
effect of oxygen: as the metallicity of the gas increases, the gas
is cooled more efficiently, the temperature drops and the line ratio
$R_3$ decreases \citep{McCall85,EvansDopita85}.

\begin{figure*}[!t]
\includegraphics[width=0.95\textwidth]{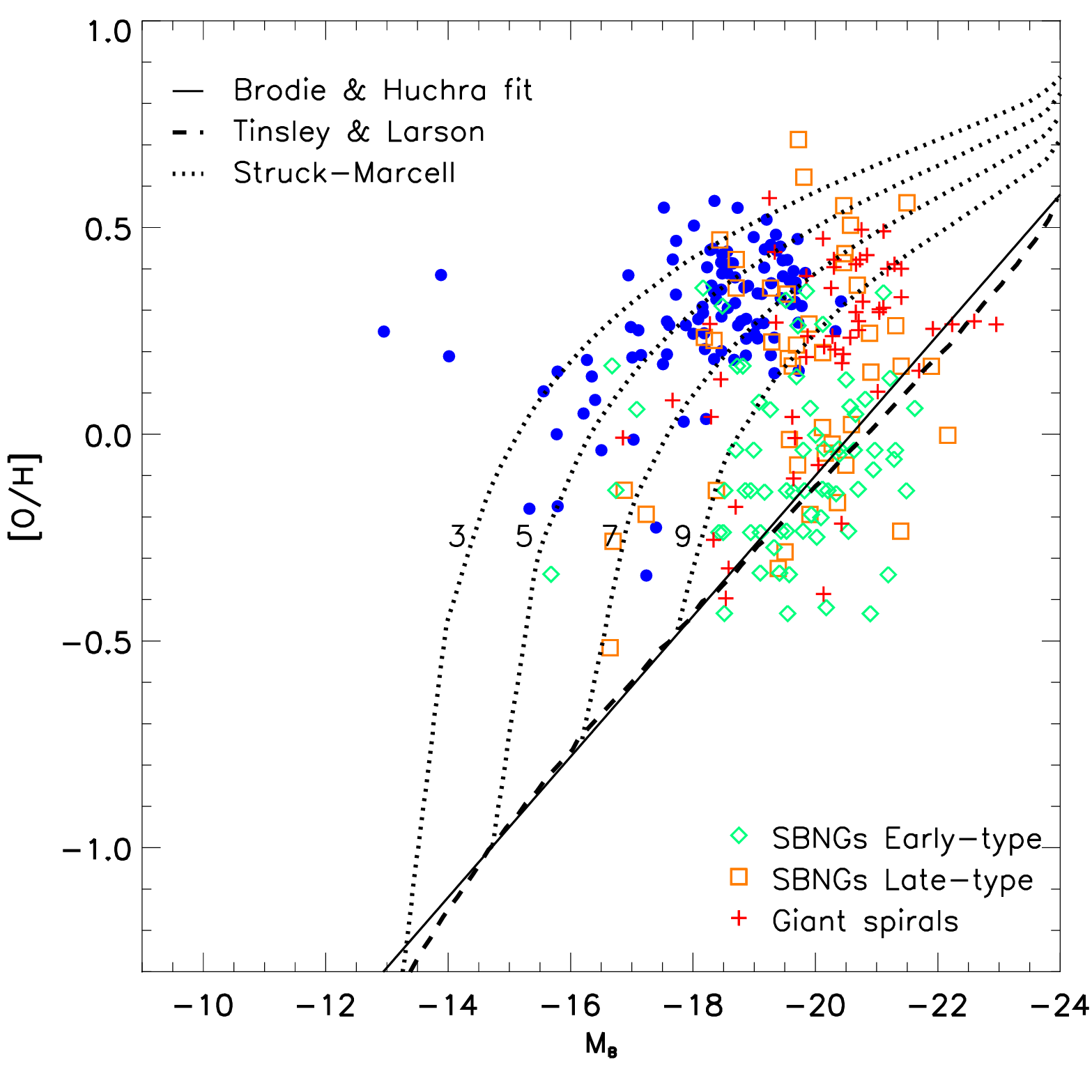}
\caption{Gas metallicities vs. absolute B magnitudes for the SFGs
(closed dots). The giant spirals sample are from \citet{ZKH94} and
the SBNGs are taken from \citep{Coz98b}. The continuous staright
line is the Brodie-Huchra relation for elliptical galaxies. Also
shown are two models of multiple mergers by \citet{TL79} and
\citet{Struck81}. The numbers indicate the approximate number of
mergers necessary in the models to reproduce the metallicities.}
\label{MBvsMetal1}
\end{figure*}

\begin{table*}[!t]\centering
  \caption{Examples of gas metallicities and absolute magnitudes in B for the
  SFGs.\\
\lowercase{(\textsc{C}omplete table available in electronic form.)}}
  \begin{tabular}{lcc}
  \hline
   Identification in SDSS&   [O/H] &  M$_{B}$          \\
                         &           &                   \\
\hline
SDSS J105809.84$-$004628.8 & 0.26 & $-18.46$  \\
SDSS J135807.05$-$002332.9 & 0.43 & $-19.68$  \\
SDSS J142223.76$-$002315.5 & 0.31 & $-17.11$  \\
SDSS J115425.04$-$021910.3 & 0.20 & $-16.35$  \\
SDSS J170128.21$+$634128.0 & 0.47 & $-18.65$  \\
SDSS J214907.29$+$002650.3 & 0.51 & $-18.28$  \\
SDSS J235106.25$+$010324.1 & 0.48 & $-17.67$  \\
SDSS J021859.64$+$001948.0 & 0.44 & $-18.69$  \\
SDSS J025154.58$+$003953.3 & 0.29 & $-19.06$  \\
SDSS J003823.71$+$150222.4 & 0.38 & $-18.69$  \\
\hline \label{TableOHSFG}
\end{tabular}
\end{table*}

The gas metallicities in the SFGs are compiled in
Table~\ref{TableOHSFG}, together with their absolute magnitudes in
B. We assume the solar metallicity is $12+ \log ($O/H$) = 8.66 \pm
0.05$ \citep{Asplund04}. The uncertainty on the metallicities is of
the order of $\pm0.2$ dex (Edmunds \& Pagel 1984). In
Figure~\ref{MBvsMetal1} we plot the gas metallicities in the SFGs
against their absolute magnitudes in B. The results are in excellent
agreement with what is expected based on their late-type
morphologies \citep{ZKH94}. The results are also in good agreement
with what was observed in Starburst Nucleus Galaxies (SBNGs): the
gas in the late-type SBNGs are more metal rich than in the
early-type SBNGs, due to their different formation processes
\citep{Coz98b}.

The usual way to determine the metallicity in the AGNs is to compare
the observed line ratios with the outputs obtained using an
ionization code like
\begin{scriptsize}CLOUDY\end{scriptsize} \citep{Ferland98}. However,
the results are usually ambiguous with more than one possible
solution \citep{Nagao06}. There maybe also other physical
mechanisms, like shocks or special ionizing structures, that can
complicate the interpretation based solely on photoionization model
results \citep{Viegas92,Cooke00}.

\begin{figure*}[!t]
\includegraphics[width=0.95\textwidth]{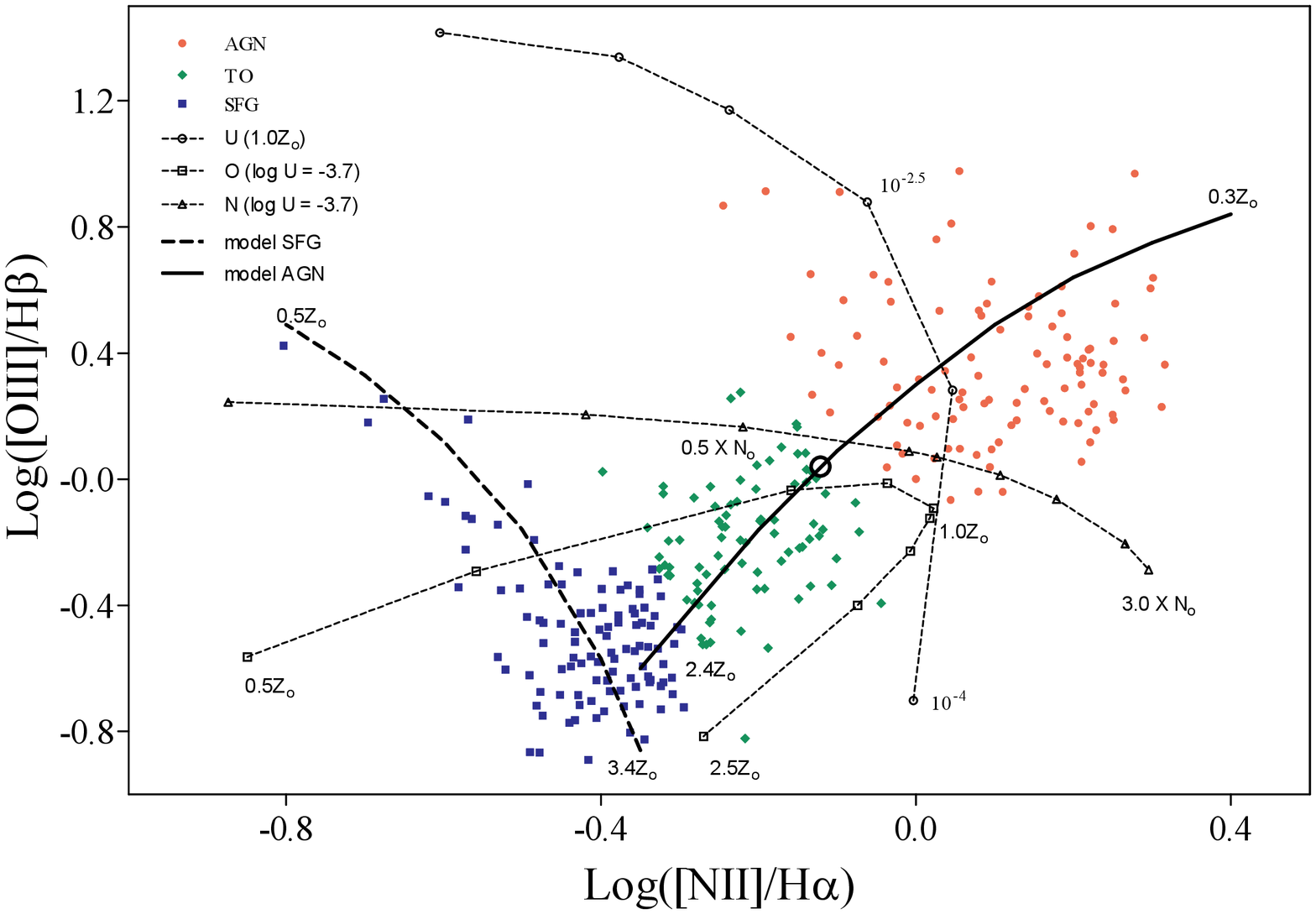}
\caption {Diagnostic diagram with three curves obtained by Bennert
et al. (2006) using {\tt CLOUDY}: U, varying ionizing parameter at
constant solar metallicity; O, varying metallicity keeping U
constant; N, varying nitrogen abundance. Also shown is our
calibration using $R_3$. The bold circle on this curve corresponds
to 1.0 Z$_\odot$.} \label{BPTcal0}
\end{figure*}

An excellent study was presented recently by \citet{Bennert06}. With
the permission of the authors, we reproduce some of their results in
Figure~\ref{BPTcal0}. The models discussed in their study apply
exactly to our NELGs. From their study we can see that the line
ratio [NII]/H$\alpha$ is extremely sensitive to the abundance of
this element. One possible cause for the increase of nitrogen
emission in the AGNs could be due to an excess of nitrogen abundance
\citep{osterbrock70,storchi89,storchi91,hamann93}. In the center of
galaxies with massive bulges and old stellar populations, we do not
expect the nitrogen abundance to follow the normal secondary
relation \citep{ZKH94,Thurston96,vanZee98,Coz99}.

In Figure~\ref{BPTcal0} we also observe that, for any value in
excess of nitrogen abundance, increasing the ionizing parameter,
$U$, while keeping the metallicity constant causes the ratio
[OIII]/H$\beta$ to rise. But, if we increase $U$ too much the ratio
[NII]/H$\alpha$ begins to decrease. In the same figure we observe
that when we lower the metallicity while keeping $U$ constant, both
line ratios increase, which is consistent with the cooling effect of
oxygen. However, if we decrease the metallicity too much, both line
ratios eventually decrease. The above behaviors are consistent with
a coupling effect between $U$ and the metallicity
\citep{EvansDopita85}: as the metallicity decreases, $U$ increases
and vice versa. Therefore, one can choose arbitrarily to change one
parameter or the other with almost the same effect.

\begin{table*}[!t]\centering
  \caption{Gas metallicities in AGNs from the literature, compared to those obtained using our calibration}
  \begin{tabular}{lccc}
  \hline
   ID & Z/Z$_\odot$ & Z/Z$_\odot$ & sources for \\
      &  (lit.) & (ours) & literature \\
\hline
Mrk 78   &  0.30     &   0.30  & (1)(2)  \\
NGC 3393 &  0.10     &   0.21  & (3)     \\
NGC 1068 &  0.30     &   0.16  & (4)     \\
NGC 4507 &  0.50     &   0.27  & (4)     \\
NGC 5135 &  0.2-0.5  &   0.43  & (4)     \\
NGC 5506 &  0.50     &   0.23  & (4)     \\
Mrk 1388 &  0.70     &   0.23  & (4)     \\
\hline \label{TableSy2Z}
\end{tabular}
\\
(1)Ramos Almeida et al. 2006, (2) Ulrich 1971, (3) Cooke et al.
2000, (4) Nagao, Maiolino \& Marconi 2006  \\

\end{table*}

In Figure~\ref{BPTcal0}, the TOs and AGNs trace a continuous
sequence where the ratios [NII]/H$\alpha$ and [OIII]/H$\beta$
increase together. According to the CLOUDY models of Bennert et al.
one way to explain this sequence would be to decrease the
metallicity gradually while increasing the excess of nitrogen
abundance. Consistent with this model, we have fitted on the TOs and
AGNs distributions an empirical relation between the ratio
[OIII]/H$\beta$ and [NII]/H$\alpha$. We have then search a
calibration that would yield the same range in metallicities for the
TOs as obtained using the model of \citet{Bennert06}, from 2.5 to 1
Z$_\odot$. As a first approximation we have found that the $R_3$
calibration for the SFGs nearly reproduces these values.
Extrapolating this calibration over the AGNs region suggests a
decrease in metallicities for these objects from 1 Z$_\odot$ to 0.3
Z$_\odot$. Based on the models of \citet{Bennert06}, the ionizing
parameter would increase from $10^{-3.5}$ to $10^{-2.5}$, which
seems in good agreement with the values found by \citet{baskin05},
and the abundance of nitrogen would not exceed 2 times the solar
value, which is a lower excess than what was suggested originally by
\citet{osterbrock70}, but still fully consistent with what is
observed in the bulges of normal galaxies
\citep{Thurston96,vanZee98,Coz99}.

Our model suggests that there is an inversion in the
metallicity-nitrogen abundance relation in the AGNs compared to that
in the SFGs: in the SFGs, the nitrogen abundance increases with the
metallicity, while in the TOs and AGNs the nitrogen abundance
increases as the metallicity decreases (consistent with an excess in
nitrogen abundance). For the TOs and AGNs the relation is:
\begin{equation}
\textrm{[O/H]} = -0.52 + (\log(\textrm{[NII]/H}\alpha) - 0.6)^2
\end{equation}
For the SFGs the relation is:
\begin{equation}
\textrm{[O/H]} = - [0.99 + 0.61 / (\log(\textrm{[NII]/H}\alpha) -
0.05)]
\end{equation}

\begin{figure*}[!t]
\includegraphics[width=0.95\textwidth]{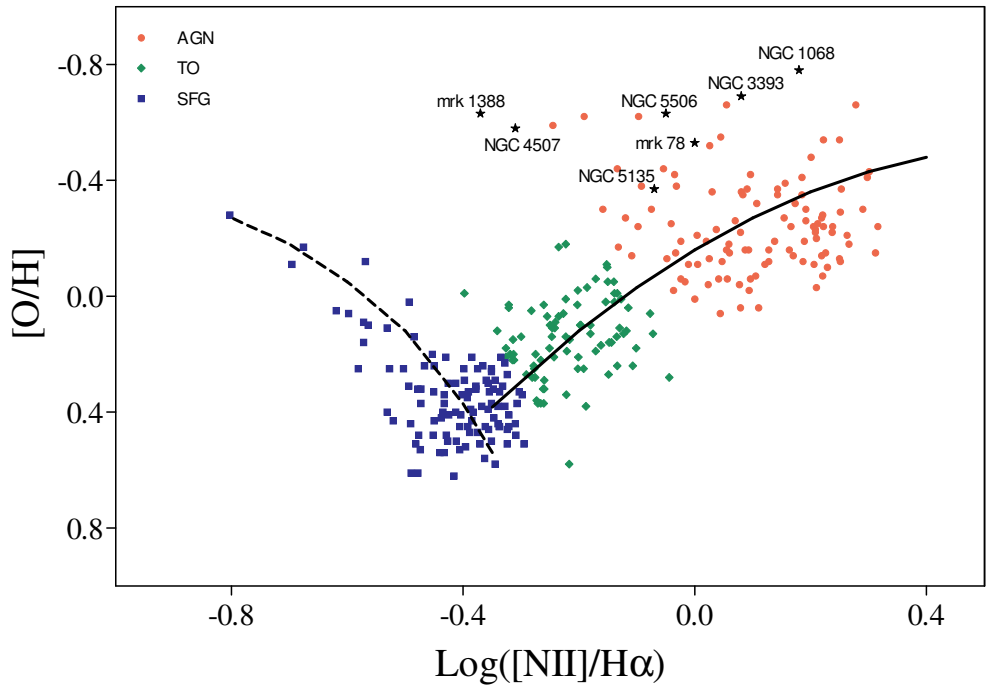}
\caption {Metallicity calibrated diagnostic diagram. The two
empirical relations (equations 3 and 4) are also traced over the
data. Also shown are Sy2 galaxies from the literature for
comparison.} \label{BPTcal1}
\end{figure*}

To test our calibration, we have searched the literature for NLAGNs
for which the metallicity of the gas was previously estimated using
\begin{scriptsize}CLOUDY\end{scriptsize}, and that could also be
evaluated by our method (with reported spectrophotometry in the
optical). We found very few examples, and all are Sy2 located in the
upper part of our diagram in Figure~\ref{BPTcal1}, where we expect
our calibration to yield the most discrepant results. The values
obtained are compiled in Table~\ref{TableSy2Z}. Only the low
metallicity solutions reported by Nagao, Maiolino \& Marconi (2006)
for their objects are consistent with our model. Surprisingly, the
differences observed are not systematic, suggesting that other
physical parameters (e.g. shocks, as we mentioned before, or special
geometries, like an ionization cone or obscuring torus) could also
be important in these galaxies.

The above comparison suggests the uncertainty on our metallicities
is of the order of 0.3 dex, increasing to as much as 0.5 dex in some
Sy2 like Mrk 1388. However, this galaxy appears like an extreme
case, for which we do not expect our calibration to apply. In
Figure~\ref{BPTcal1}, most of the AGNs and TOs in our sample fall in
a different regime than the AGNs from the literature used for the
test (that is, the AGN in our sample fall in a different part of the
diagnostic diagram, and they are clearly tracing a continuous
relation). A better estimate of the uncertainty on the gas
metallicities related with our method may be $\sim0.2$ dex, which is
consistent with the variance of our fitted relation.

\begin{table*}[!t]\centering
  \caption{Examples of gas metallicities and absolute magnitudes in B for the
  AGNs.\\
\lowercase{(\textsc{C}omplete table available in electronic form.)}}
  \begin{tabular}{lcc}
  \hline
   Identification in SDSS &   [O/H] &  M$_{B}$          \\
                          &           &                   \\
\hline
SDSS J113903.33$-$001221.6&$-0.15$& $-18.06$  \\
SDSS J150654.85$+$001110.8&$-0.16$& $-19.52$  \\
SDSS J113423.32$-$023145.5&$-0.24$& $-19.75$  \\
SDSS J172613.73$+$620858.1&$-0.19$& $-18.85$  \\
SDSS J173044.85$+$562107.1&$-0.35$& $-19.25$  \\
SDSS J025017.75$-$083548.5&$-0.20$& $-17.84$  \\
SDSS J032501.68$-$054444.8&$-0.06$& $-18.25$  \\
SDSS J142757.71$+$625609.3&$-0.22$& $-19.46$  \\
SDSS J114743.68$+$014934.3&$-0.66$& $-19.12$  \\
SDSS J133548.24$+$025956.1&$-0.36$& $-19.28$  \\
\hline \label{TableOHAGN}
\end{tabular}
\end{table*}

\begin{table*}[!t]\centering
  \caption{Examples of gas metallicities and absolute magnitudes in B for the
  TOs.\\
\lowercase{(\textsc{C}omplete table available in electronic form.)}}
  \begin{tabular}{lcc}
  \hline
   Identification in SDSS &   [O/H] &  M$_{B}$          \\
                          &           &                   \\
\hline
SDSS J122353.98$-032634.4$&$\ \ 0.16$& $-17.40$  \\
SDSS J124428.12$-030018.8$&$\ \ 0.14$& $-19.10$  \\
SDSS J224424.36$-000943.5$&$\ \ 0.24$& $-19.44$  \\
SDSS J020540.31$-004141.4$&$\ \ 0.21$& $-19.37$  \\
SDSS J031347.83$+004139.7$&$\ \ 0.10$& $-19.27$  \\
SDSS J032406.50$-010328.2$&$\ \ 0.03$& $-18.49$  \\
SDSS J012853.25$+134737.6$&$\ \ 0.02$& $-18.95$  \\
SDSS J030848.32$-070226.1$&$\ \ 0.28$& $-19.53$  \\
SDSS J033358.81$-070826.6$&$\ \ 0.10$& $-18.82$  \\
SDSS J090513.20$-002947.8$&$-0.01$& $-17.82$  \\
\hline \label{TableOHTO}
\end{tabular}
\end{table*}

\begin{figure*}[!t]
\includegraphics[width=0.95\textwidth]{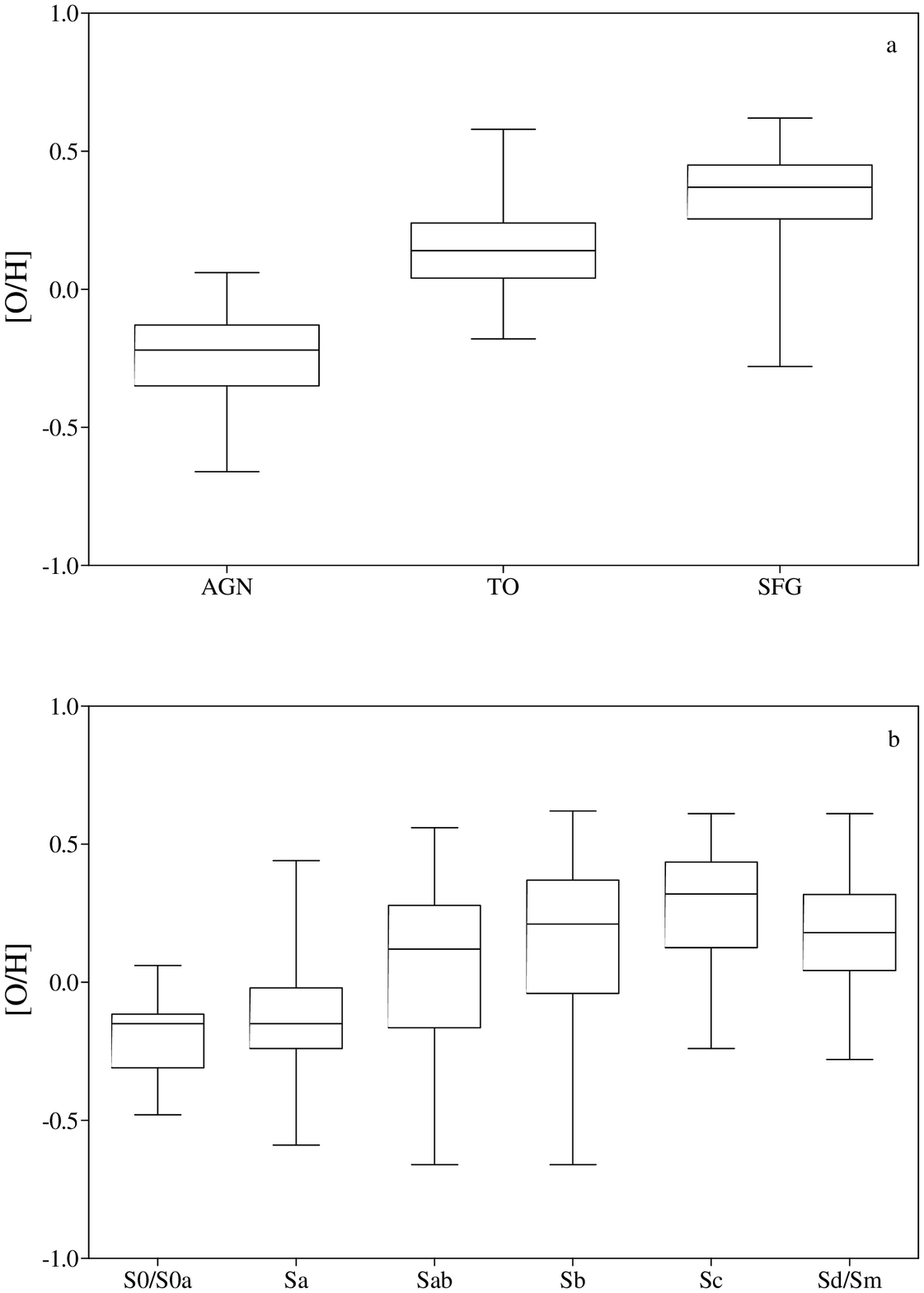}
\caption{Box-whisker plots for the gas metallicity varying in
galaxies having a) different activity types, and b) different
morphologies.} \label{BW_ZvsT}
\end{figure*}

The gas metallicities as deduced from our calibration together with
the luminosities in B are presented in Table~\ref{TableOHAGN} for
the AGNs, and in Table~\ref{TableOHTO} for the TOs. In
Figure~\ref{BW_ZvsT} we show the box-whisker plots for the gas
metallicities in galaxies having different activity types and
different morphologies. The AGNs and TOs have lower gas
metallicities than the SFGs and we find an excellent correlation
with the morphological type of the galaxies, the metallicities
increasing in the later types. The statistical tests
(Table~\ref{test1} and Table~\ref{test2} in the appendix) confirm
the differences observed at a level of confidence of 99\%.

\begin{figure*}[!t]
\includegraphics[width=0.95\textwidth]{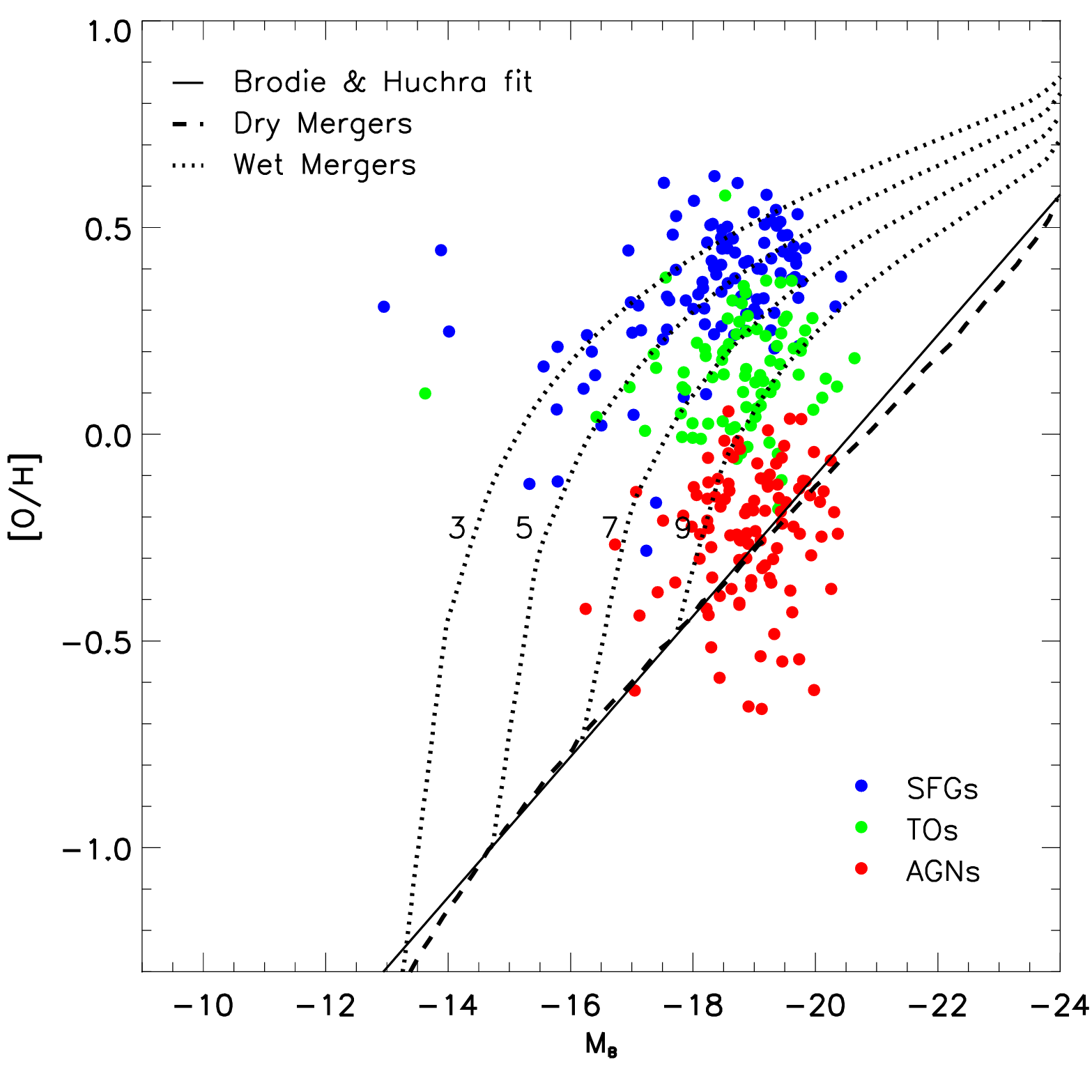}
\caption{Gas metallicities vs. absolute B magnitudes for all the
isolated NELGs. The multiple merger model of \citet{TL79} is
identified as the dry merger scenario and the \citet{Struck81} model
is qualified as the wet merger scenario (see explanations in our
discussion section). } \label{MBvsMetal2}
\end{figure*}

In Figure~\ref{MBvsMetal2} we compare the gas metallicities of the
TOs and AGNs with those of the SFGs. The AGNs seem to follow the
Brodie-Huchra mass-metallicity relation for elliptical and
bulge-dominated galaxies \citep{ZKH94,Coz98b}. The TOs, on the other
hand, with their intermediate morphologies, show also metallicities
which are intermediate between those of the SFGs and AGNs. The
differences in metallicities observed between the SFGs, TOs and AGNs
are in excellent agreement with the differences in mean ages of the
stellar populations and bulge masses. All these parameters are
consistent with higher astration rates for the AGNs as compared to
the SFGs \citep{Sandage86}.

\section{Discussion}

Our study suggests that the formation of a SMBH in the center of a
galaxy is tightly connected with the formation process of its bulge
\citep{haring04,peterson05,Gultekin09}. Some authors even suggested
that the formation of the bulge is a self-regulated process with
strong feedback from the SMBH growing in its center (see Younger et
al. 2008 and references therein). Our observations may support such
interpretations.

\begin{figure*}[!t]
\includegraphics[width=0.95\textwidth]{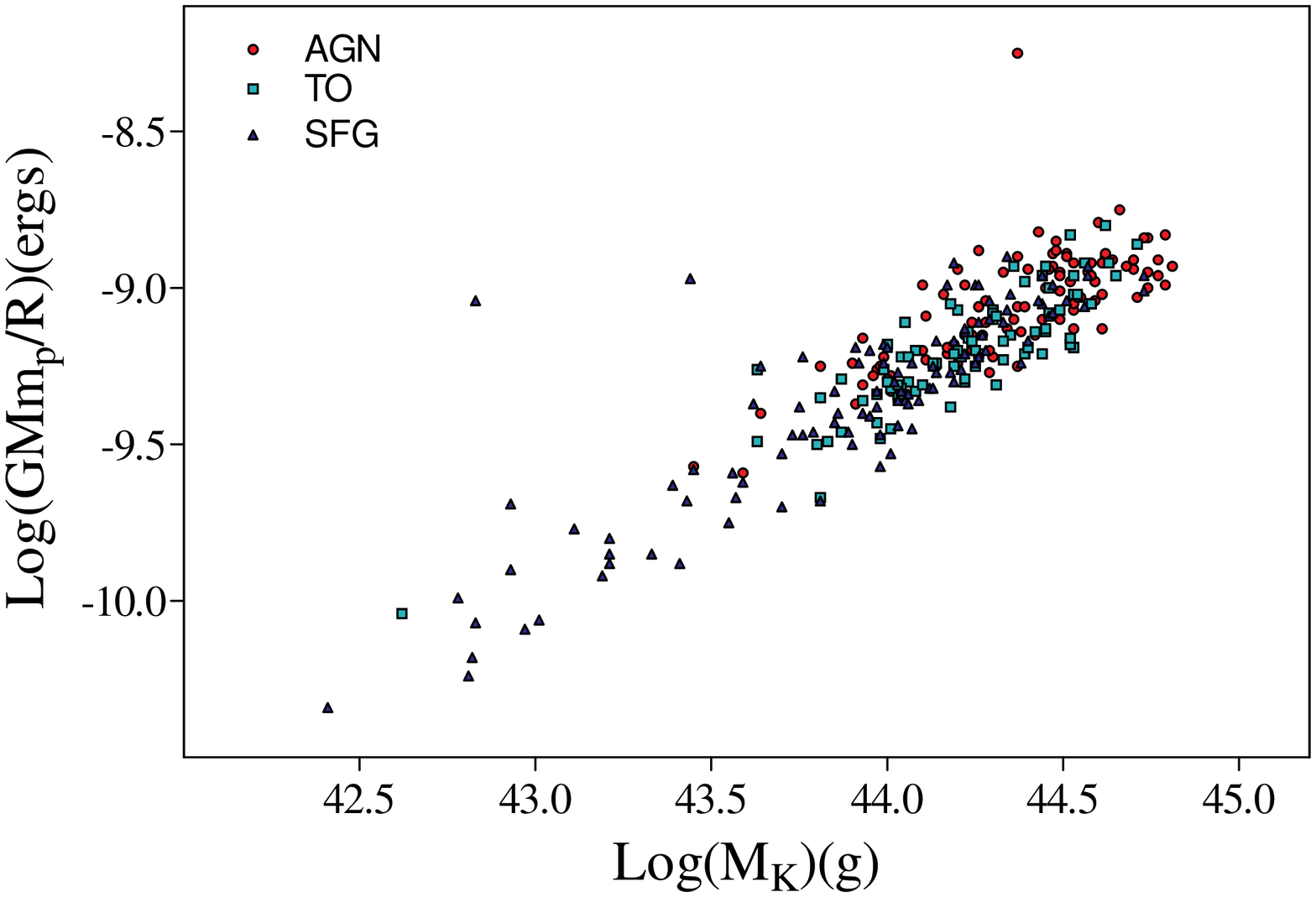}
\caption{Gravitational potential energy per baryon of the galaxies
in our sample.} \label{EGgal}
\end{figure*}

According to \citet{Aller07} the gravitational binding energy is one
key factor explaining the relation between the SMBH and the bulge
mass. The gravitational binding energy is the negative of the
gravitational potential energy. For a system of $N$\ particles, with
an approximated mass $M=N m_p$, where $m_p$ is the mass of a proton,
the binding energy per baryon is equal to:
\begin{equation}
 U/N= G M m_p/R
\end{equation}
where $G$ is the gravitational constant and $R$ is the radius of the
object. By definition, a SMBH represent a huge mass, which
represents maybe only 1\% to 0.1\% of the the bulge mass, but which
is concentrated in an extremely small region of space at the center
of mass of the galaxy. The formation of such a highly
gravitationally bound object implies a significant increase in
binding energy of the host galaxy itself.

\begin{figure*}[!t]
\includegraphics[width=0.95\textwidth]{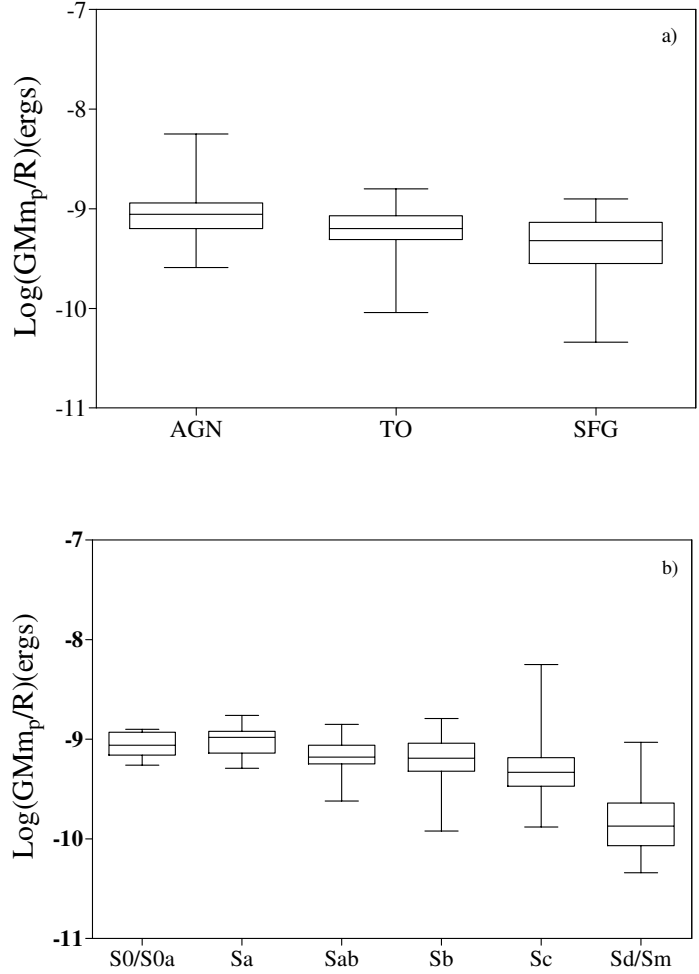}
\caption{Box-whisker plot of gravitational potential energy per
baryon of the isolated NELGs, a) with different activity types, b)
with different morphologies.} \label{BWEGgal}
\end{figure*}

The above description gives us a new test for the NLAGNs in our
sample. If these galaxies host a SMBH in their center, we would thus
expect them to show relatively high gravitational binding energies
compared to, for example, the SFGs. To verify this assumption, we
have used the masses as determined from the K magnitudes, $M_K$, and
the Petrosian radii, $R_{90\%}$, reported in Table~2, to calculate
the gravitational binding energy per baryon of all the galaxies in
our sample. As expected, we observe in Figure~\ref{EGgal} that the
AGNs and TOs have higher gravitational binding energies than the
SFGs. This trend is better seen in Figure~\ref{BWEGgal}a, which
shows the box-whisker plots for the gravitational binding energies.
The trend is also confirmed to be statistically significant in
Table~\ref{test1} of the appendix.

The variation of gravitational binding energies with galaxy
morphologies is shown in Figure~\ref{BWEGgal}b, and the trend is
statistically tested in Table~\ref{test2} of the appendix. Galaxies
with massive bulges, S0, S0a, and Sa, have comparable gravitational
binding energies, which are significantly higher than for galaxies
having less massive bulges, Sab and later. These results are in good
agreement with the hypothesis of a SMBH in the NLAGNs (and possibly
also in the TOs).

\subsection{Masses of SMBHs vs. accretion rates}

Our analysis of the NLAGNs in our sample are consistent with the
standard AGN interpretation: this phenomenon is the product of the
accretion of matter onto a SMBH that developed at the same time as
the massive bulge of the galaxy. According to this interpretation,
the nearby NLAGNs in our sample are possibly scaled down (and/or
power down) versions of quasars and broad-line AGNs (BLAGNs). One
good example is our own galaxy, with an intermediate Sb or SBb
morphology \citep{binney98}, where evidence was found in its center
for both a SBMH, with a mass of the order of 3 or $\sim4\times10^{6}$ M$_{\odot}$
\citep{GT87,GMB00,SME09}, and of recent star formation episodes
\citep{Figer04}, which suggests she could be classified as a LINER
or a TO by outside observers.

\begin{figure*}[!t]
\includegraphics[width=0.95\textwidth]{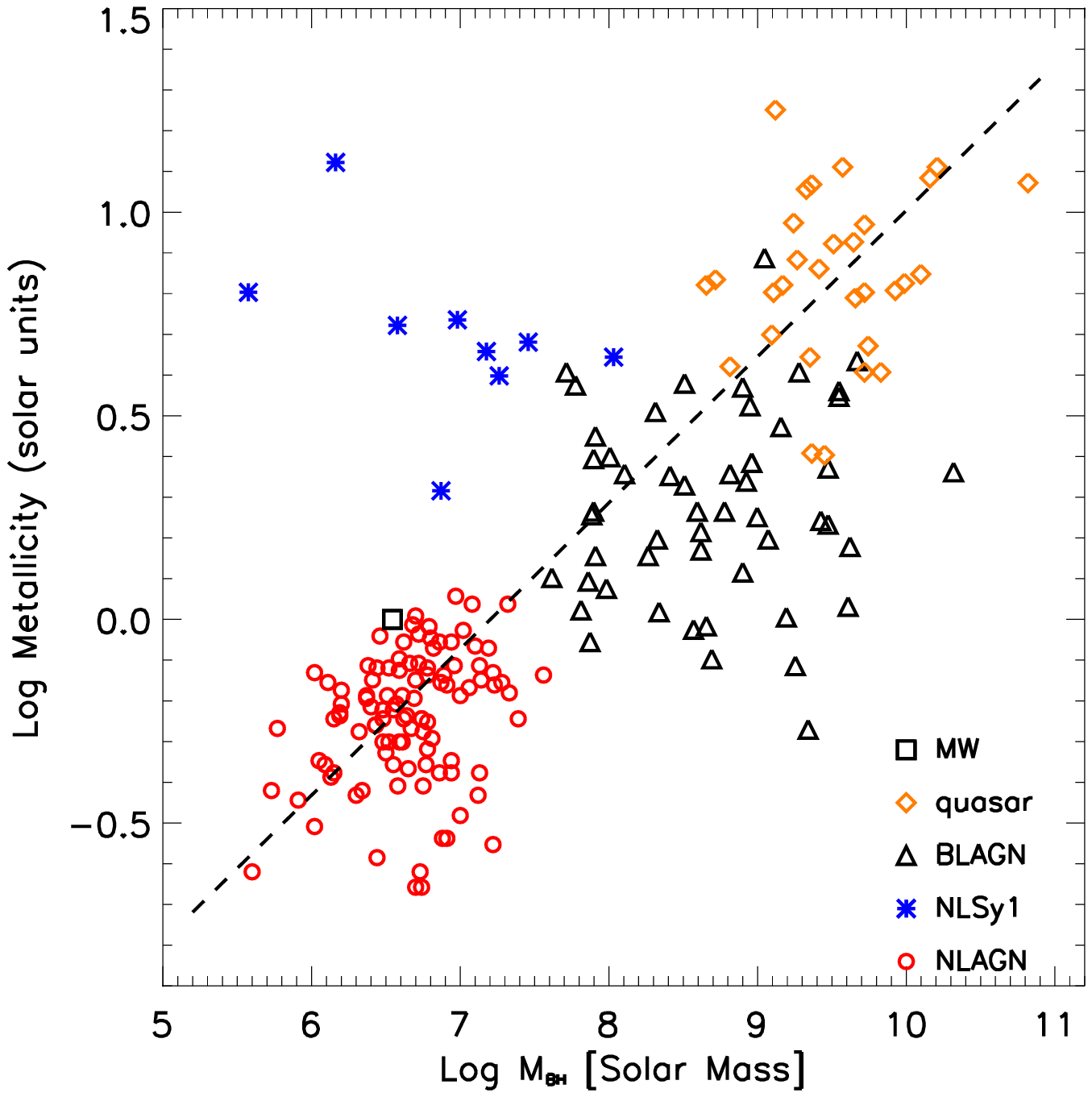}
\caption{Metallicities vs. Black Hole mass. The dashed line is our
linear fitted correlation (excluding the NLSy1). The metallicity for
the Milky Way is that of the Sun, which is now in better agreement
with those of stars in its immediate neighborhood (Asplund et al.
2004).} \label{OHvsMBH}
\end{figure*}

In \citet{haring04} a strong correlation was encountered between the
bulge mass and black hole mass (see also G\"ultekin et al. 2009 and
reference therein). In Shemmer et al. (2004) and Matsuoka et al.
(2011), a strong correlation was also found between the black hole
mass and the gas metallicity. These two correlations are fully
consistent with our observations: while the formation of the SMBH
follows (or self-regulate) the formation of the bulge, the gas
metallicity, being a product of the evolution of the stars, also
depends on the bulge formation through typically high astration
rates \citep{Sandage86}. Based on these observations, we may
therefore expect the NLAGNs to follow the same correlations as the
quasars and BLAGNs.

\begin{figure*}[!t]
\includegraphics[width=0.95\textwidth]{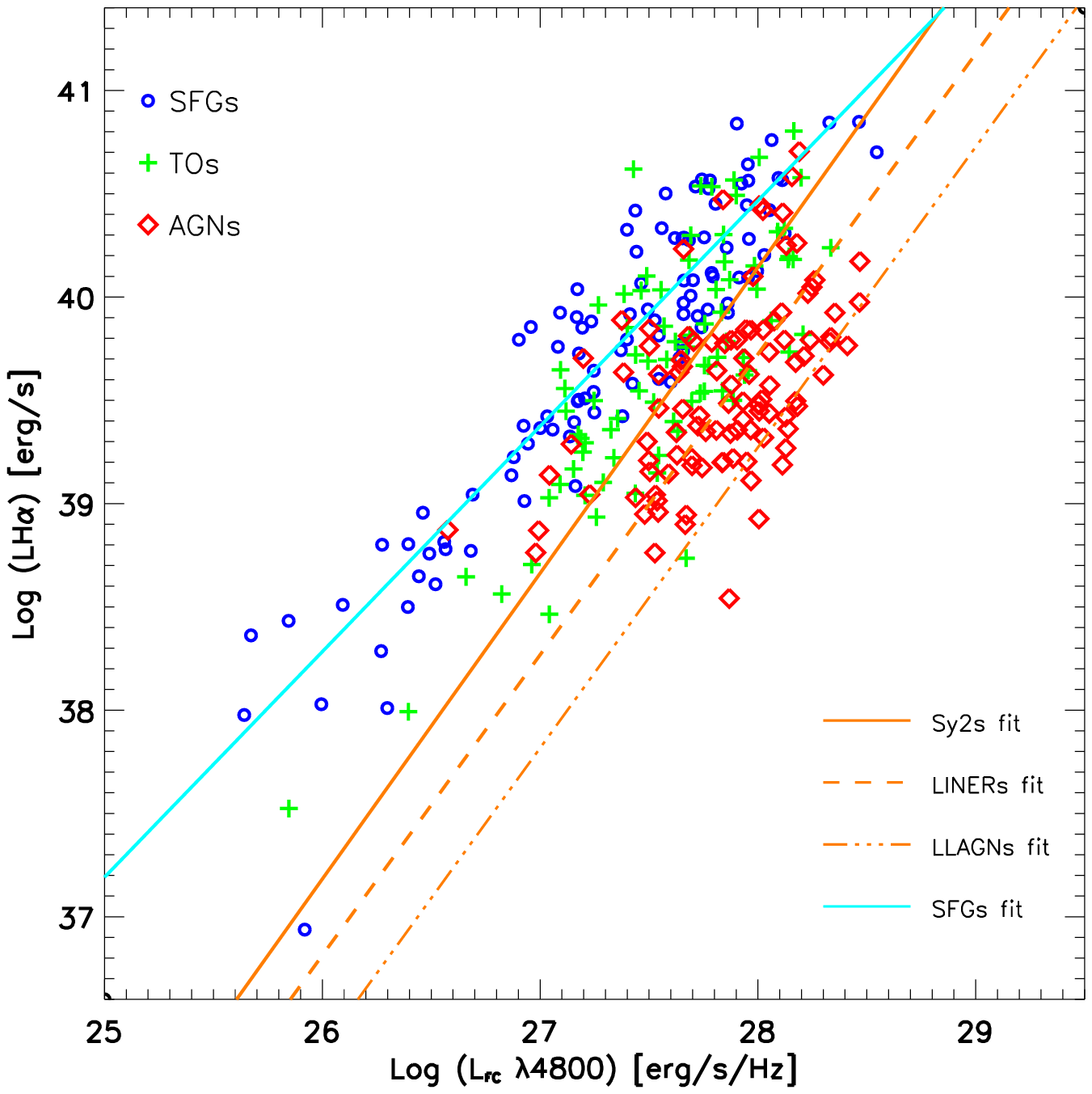}
\caption{Relation between ionization luminosity (H$\alpha$) and
continuum luminosity (at 4800 \AA). The different linear relations
were determined by \citet{TP11} using  a sample of 318486 SDSS
galaxies. The SFGs follow a linear relation $L_{H\alpha}\propto
L^{1.09\pm0.03}_{4800}$, while the NLAGNs follow a steeper power law
relation $L_{H\alpha}\propto L^{1.45\pm0.03}_{4800}$.}
\label{powerlaw}
\end{figure*}

To calculate the black hole masses, M$_{BH}$, as reported in
column~3 of Table~\ref{MASS}, we used the relation between the bulge
mass and black hole mass as determined by \citet{haring04}. For the
isolated NLAGNs we find a median value of $4.8\times10^{6}$
M$_{\odot}$, which is two to three orders below the value found in
BLAGNs, but in good agreement with the black hole mass found in the
center of our galaxy \citep{GT87,GMB00,SME09}.

In Figure~\ref{OHvsMBH} we compare the gas metallicities and black
hole masses for the NLAGNs in our sample with those measured in
quasars and BLAGNs \citep{shemmer04}. The metallicities measured by
\citet{shemmer04} was transformed to the unity used in our study  by
\citet{Neri-Larios11}. The NLAGNs seem to continue the linear
correlation found for the quasars and BLAGNs in the lower
metallicity regime. A linear fit with a correlation coefficient of
$r_{Pearson} = 0.77$ and  $r_{Spearman}=0.83$, both with chance
probability $P(r_{Pearson})$ and $P(r_{Spearman})$ practically equal
to zero, suggests the metallicity increases with the black hole mass
as:
\begin{equation}
 \log(\textrm{[O/H]}) = -2.56 + 0.3590\times \log(M_{BH})
\end{equation}

\begin{figure*}[!t]
\includegraphics[width=0.95\textwidth]{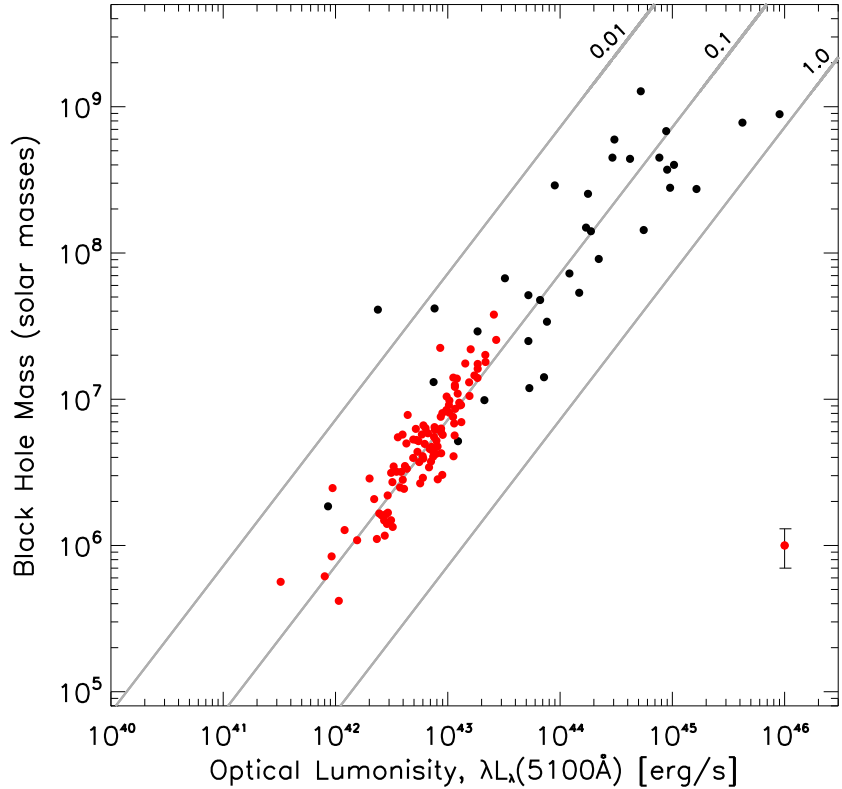}
\caption{Black hole mass as a function of continuum luminosity at
5100 \AA\ for the NLAGNs compared to the BLAGNs studies in Peterson
et al. (2005). The typical uncertainty for the NLAGNs is indicated
by the red dot with error bars.} \label{accretion}
\end{figure*}

In \citet{TP11} it was shown that all the different kinds of NLAGNs
(Seyfert 2, LINERS and LLAGNs) follow the same power law between the
luminosity in H$\alpha$ and luminosity at 4800 \AA\
\citep{osterbrock89}, while the SFGs follow a different, less steep
linear relation. In Figure~\ref{powerlaw} we verified that this also
applies to the isolated NLAGNs in our sample. One can note also the
intermediate position of the TOs, consistent with being a mixture of
SFGs and AGNs. This result suggests that we can use the black hole
masses in conjunction with the luminosities at 5100\AA\ to estimate
their accretion rates, as was done by \citet{peterson05}.

In Figure~\ref{accretion}, we follow the method of
\citet{peterson05} to estimate the accretion rates of the NLAGNs in
our sample. The SMBHs in the NLAGNs seem to accrete matter at
relatively high rates of 0.1 times the Eddington limit. The isolated
NLAGNs are consistent with scaled-down versions of quasars and
BLAGNs, not powered-down versions: the lower luminosity is a result
of smaller-mass black holes, not lower accretion rates.

\begin{figure*}[!t]
\includegraphics[width=0.95\textwidth]{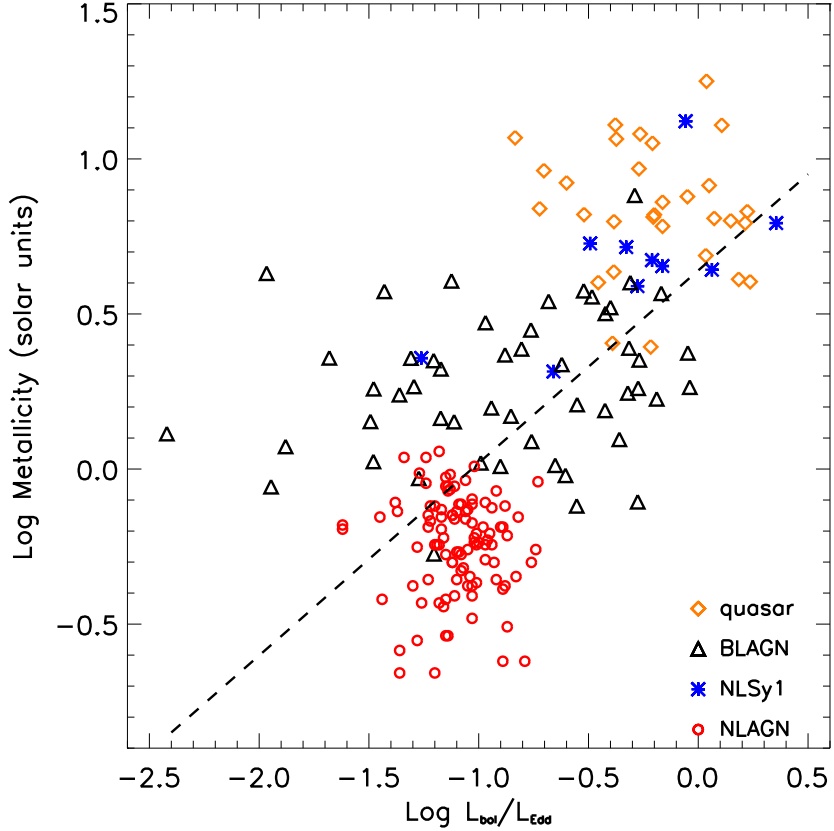}
\caption{Relation between metallicity and accretion rate. The NLAGNs
are compared to the quasars and BLAGNs as studied by
\citep{shemmer04}. The linear relation was fitted on all the
galaxies (including the NLSy1)}. \label{OHvsAccretion}
\end{figure*}

In Figure~\ref{OHvsAccretion} we compare the gas metallicities and
accretion rates of the NLAGNs in our sample with those found in
quasars and BLAGNs, as determined in \citet{Neri-Larios11} for the
sample studied by \citet{shemmer04}. Again, the NLAGNs seem to
continue the linear correlation found for the quasars and BLAGNs in
the lower metallicity regime. A linear fit with a correlation
coefficient $r_{Pearson} = r_{Spearman}= 0.66$ with both chance
probabilities practically equal to zero, suggests the metallicity
increases with the accretion rate as:
\begin{equation}
 \log(\textrm{[O/H]}) = 0.77 + 0.66\times \log(L_{Bol}/L_{Edd})
\end{equation}

Although our results are fully consistent with the standard
interpretation of NLAGNs as scaled-down versions of quasars and
BLAGNs, we must state that there is an unfortunate uncertainty on
the black hole masses we have determined. The fact is that there is
no consensus in the literature about what is the correct relation to
use. For practical reason we have used the relation proposed by
\citet{haring04} between the bulge mass and the black hole mass.
However, in  \citet{Gultekin09} the relation proposed between the
black hole mass and velocity dispersion of the bulge yields masses
which are ten times higher. The problem lies with what
\citet{haring04} called the ``black hole sphere of influence''.
These authors argued that the black hole representing only a small
fraction of the bulge mass (1\% to 0.1\%) cannot be responsible for
the full value of the velocity dispersion, which explains why they
apllies a correction and obtain ten times lower black hole mass
values. In \citet{Gultekin09}, this correction is simply rejected as
non valid, without further arguments.

If we adopt the relation of \citet{Gultekin09}, then the SMBH in the
NLAGNs would be ten times more massive. In Figure~\ref{OHvsMBH}, the
relation between the metallicity and black hole mass would be
slightly steeper and the NLAGNs would differ significantly from the
Milky Way. In Figure~\ref{accretion} the NLAGNs would not be
scaled-down versions of BLAGNs but rather powered-down versions,
that is, the SMBH of the NLAGNs have masses comparable to those
found in the BLAGNs but they would now be accreting at only 0.01
times the Eddington limit. Finally in Figure~\ref{OHvsAccretion} the
NLAGNs would fall farther to the left, but would still be in good
agreement with the metallicity vs. accretion rate relation suggested
by \citet{shemmer04}.

\section{Summary and conclusion}

We have constructed a sample of galaxies that have formed in low
galactic density environments and have evolved in relative
isolation. All these galaxies show a spiral disk and some kind of
nuclear activity. Using a standard diagnostic diagram we have
established that as much as 64\% of these galaxies are classified as
NLAGNs or TOs.

We have established a strong connection between the AGN phenomenon
and the mass of the bulge, but also found a strong correlation with
the morphological type of the galaxies. It suggests that the AGN
phenomenon is a ``normal'' occurrence related with the formation
process of massive bulges in early-type spiral galaxies.

Consistent with this interpretation, our analysis suggests the AGNs
and TOs have experienced during their formation higher astration
rates than the SFGs--transforming their gas into stars more
efficiently \citep{Sandage86}. Evidence favoring higher astration
rates in AGNs than in SFGs are: 1) their older stellar populations;
2) their lower oxygen abundance and a possible excess of nitrogen.
We also found the NLAGNs to have higher binding energies than the
SFGs, suggesting they host a SMBH in their center.

Our results are in good agreement with the standard interpretation
of NLAGNs, either as scaled-down (lower BH mass but same accretion
rates) or powered-down (same BH mass but lower accretion rates)
versions of quasars and BLAGNs. The NLAGNs also seem to follow the
same relation between the BH mass, the accretion rate and gas
metallicity as the quasar and BLAGNs.

\section*{Acknowledgments}
We dedicate this article to the memory of Gabriel Garc\'{\i}a
Ru\'{\i}z, night assistant and friend at the Observatorio
Astron\'omico Nacional, San Pedro M\'artir, M\'exico, who died
tragically while on duty at the end of 2010. The authors would also
like to thank N. Bennert and S. Komossa for making their results of
CLOUDY models available to us, and O. Shemmer for furnishing the
data for the quasars and BLAGNs. T-P acknowledges PROMEP for support
grant 103.5-10-4684 and I. P.-F. acknowledges a postdoctoral
fellowship from CONACyT, M\'exico No. 145727. This publication makes
use of data products from the Two Micron All Sky Survey, which is a
joint project of the University of Massachusetts and the Infrared
Processing and Analysis Center/California Institute of Technology,
funded by the National Aeronautics and Space Administration and the
National Science Foundation. The SDSS is managed by the
Astrophysical Research Consortium (ARC) for the Participating
Institutions. The Participating Institutions are: the American
Museum of Natural History, Astrophysical Institute Potsdam,
University of Basel, University of Cambridge (Cambridge University),
Case Western Reserve University, the University of Chicago, the
Fermi National Accelerator Laboratory (Fermilab), the Institute for
Advanced Study, the Japan Participation Group, the Johns Hopkins
University, the Joint Institute for Nuclear Astrophysics, the Kavli
Institute for Particle Astrophysics and Cosmology, the Korean
Scientist Group, the Los Alamos National Laboratory, the
Max-Planck-Institute for Astronomy (MPIA), the Max-Planck-Institute
for Astrophysics (MPA), the New Mexico State University, the Ohio
State University, the University of Pittsburgh, University of
Portsmouth, Princeton University, the United States Naval
Observatory, and the University of Washington.

\appendix

\section{Results of statistical test}

The statistical tests used in this study are the non-parametric
Kruskal-Wallis test (KW), together with the Dunn's post tests. The
KW test compares the medians in three or more unmatched groups of
data and the post tests do multiple one-to-one comparisons. The P
values reported in Table~\ref{test1} and Table~\ref{test2} are
codified in the following way: non significant difference, ns,
significant,
**  for P$<0.01$, highly significant, ***
for P$<0.001$.

In Table~\ref{test1}, comparing the bulge mass in galaxies having
different activity types, the post tests indicate that all three
samples have different medians. In group~1, no difference is
encountered while in group~2 only the AGNs differ from the SFGs.

In Table~\ref{test2}, we compare the bulge mass in galaxies having
different morphologies, the post tests find highly significant
differences between the most separated morphological classes. No
differences are observed between the S0, S0a, Sa and Sab, the Sab
and Sb, and the Sc and Sd. These tests suggest that the relation
between the bulge mass and activity type is a by-product of the
morphology, consistent with the results for the diagnostic diagrams.

In Table~\ref{test1}, we verify that the total mass does not vary as
strongly as the bulge mass in galaxies having different activity
types. The same is true for the morphologies in Table~\ref{test2}.
Therefore, the trend observed for the AGN activity is more
qualitative than quantitative: it is not the mass of the galaxy that
counts, but the fraction of mass that is in the form of a bulge.

In Table~\ref{test1} we find significant differences in
metallicities, the AGNs being less metal rich than the SFGs, and the
TOs being intermediate. We also find a significant difference in
metallicity between the early and late-type galaxies in
Table~\ref{test2}, early-type galaxies being metal poor compared to
the late-type ones. These results are in good agreement with the
difference in mean stellar population ages.

In Table~\ref{test1} we find a strong relation between the
gravitational binding energy and the AGN activity. This applies for
galaxies having more massive bulges, Table~\ref{test2}. Since we
find no such relation for the masses, these results are consistent
with a strong dependence of the AGN phenomenon with the
gravitational binding energy: the higher the gravitational binding
energy, the more higher the bulge mass and the higher the
probability to find an AGN.

All these results favor an explanation in terms of astration rates:
high astration rates produce massive bulges which increases the
probability to find an AGN.

\begin{table*}[!t]\centering
  \small
  \begin{changemargin}{-1.5cm}{-1.5cm}
\caption{Dunn's post tests for activity types vs. bulge mass, total
mass, metallicity and gravitational binding energy}
  \begin{tabular}{lcccccccccccccc}
\hline
          &\multicolumn{2}{c}{$M_{Bulge}$}       &\multicolumn{2}{c}{group 1} &\multicolumn{2}{c}{group 2} &\multicolumn{2}{c}{$M_{B}$} &\multicolumn{2}{c}{$M_{K}$} &\multicolumn{2}{c}{[O/H]} &\multicolumn{2}{c}{U/N}\\
          & AGN  & TO    & AGN  & TO                  & AGN & TO                   & AGN  & TO                  & AGN  & TO                  & AGN  & TO & AGN  & TO                 \\
\hline
AGN       &      &       &      &                     &      &                     &      &                     &      &                     &      &        &      &                  \\
TO        & ***  &       &  ns  &                     &  ns  &                     & ns   &                     & ns   &                     & ***  &        & ***  &                  \\
SFG       & ***  &  ***  &  ns  & ns                  &  *** & ns                  & ***  &    **               & ***  &  ***                & ***  &  ***   & ***  &    **            \\

\hline
 \label{test1}
\end{tabular}
\end{changemargin}
\end{table*}

\begin{table*}[!t]\centering
  \small
  \begin{changemargin}{-1.5cm}{-1.5cm}
\caption{Dunn's post tests for morphology vs. bulge mass, total
mass, metallicity and gravitational binding energy}
  \begin{tabular}{lccccccccccccccc}
\hline    &\multicolumn{5}{c}{$M_{Bulge}$}&\multicolumn{5}{c}{[O/H]}     &\multicolumn{5}{c}{U/N}        \\
          & S0/S0a & Sa & Sab & Sb & Sc   & S0/S0a & Sa  & Sab & Sb &    & S0/S0a & Sa  & Sab & Sb & Sc  \\
\hline
Sa        & ns     &    &     &    &      & ns     &     &     &    &    & ns     &     &     &    &     \\
Sab       & ns     & ns &     &    &      & ns     & **  &     &    &    & ns     & *** &     &    &     \\
Sb        & **     & ***& ns  &    &      & **     & *** & ns  &    &    & ns     & *** & ns  &    &     \\
Sc        & ***    & ***& *** & ** &      & ***    & *** & *** & ns &    & ***    & *** & **  & ns &     \\
Sd        & ***    & ***& *** & ***& ns   &        &     &     &    &    & ***    & *** & *** & ***& ns  \\
\hline \hline
         &\multicolumn{5}{c}{$M_{B}$} &\multicolumn{5}{c}{$M_{K}$}    &&&&  \\
         & S0/S0a & Sa & Sab & Sb & Sc & S0/S0a & Sa & Sab & Sb & Sc  &&&&  \\
\hline
Sa       & ns     &    &     &    &    & ns     &    &     &    &    &&&&   \\
Sab      & ns     & ns &     &    &    & ns     & ns &     &    &    &&&&   \\
Sb       & ns     & ns & ns  &    &    & ns     & ** & ns  &    &    &&&&   \\
Sc       & **     & ** & ns  & ***&    & ns     & ***& ns  & ns &    &&&&   \\
Sd       & ***    & ***& *** & ***& ** & ***    & ***& *** & ***& ** &&&&   \\
\hline \label{test2}
\end{tabular}
\end{changemargin}
\end{table*}

\end{document}